\pdfoutput=1
\documentclass[twocolumn]{aastex63}
\usepackage{subfigure}
\usepackage{amsmath}

\usepackage{soul}
\usepackage{lineno}

\newcommand{\x}{X-ray}

\newcommand{\chandra}{\textit{Chandra}}

\newcommand{\swift}{\textit{Swift}}

\newcommand{\xmm}{\textit{XMM-Newton}}

\newcommand{\rxte}{{\it RXTE}}
\newcommand{\nustar}{\textit{NuSTAR}}

\newcommand{\sax}{{\it BeppoSAX}}

\newcommand{\integral}{\textit{INTEGRAL}}

\newcommand{\heao}{{\it HEAO 1}}

\newcommand{\exi}{\begin{equation}}
\newcommand{\exo}{\end{equation}}

\def\la{\mathrel{\hbox{\rlap{\hbox{\lower4pt\hbox{$\sim$}}}\hbox{$<$}}}}
\def\ga{\mathrel{\hbox{\rlap{\hbox{\lower4pt\hbox{$\sim$}}}\hbox{$>$}}}}
\def\lesssim{\mathrel{\spose{\lower 3pt\hbox{$\mathchar"218$}}
     \raise 2.0pt\hbox{$\mathchar"13C$}}}
\def\gtrsim{\mathrel{\spose{\lower 3pt\hbox{$\mathchar"218$}}
     \raise 2.0pt\hbox{$\mathchar"13E$}}}
\def\arcdeg{\hbox{$^\circ$}}

\definecolor{charcoal}{rgb}{0.21, 0.27, 0.31}
\definecolor{coolblack}{rgb}{0.0, 0.18, 0.39}
\definecolor{darkblue}{rgb}{0.0, 0.0, 0.55}

\received{November 17, 2023}
\revised{April 10, 2023}
\accepted{\today}
\submitjournal{Apj}

\shorttitle{\nustar\ CXB Measurement}
\shortauthors{Rossland et al.}
\graphicspath{{./}{figures/}}

\begin{document}

\title{Measuring the Cosmic X-ray Background in 3-20~KeV with Straylight from \nustar

\footnote{Released on April 13, 2023}}

\email{coimhead1643@gmail.com}

\author{Steven Rossland}
\affiliation{University of Utah, 
115 1400 E, 
Salt Lake City, UT 84112, USA}

\author{Daniel R.\ Wik}
\affiliation{University of Utah, 
115 1400 E,  
Salt Lake City, UT 84112, USA}


\author{Brian Grefenstette}
\affiliation{Cahill Center for Astronomy and Astrophysics, 
California Institute of Technology, 
Pasadena, CA 91124, USA}

\author{Nico Cappelluti}
\affiliation{Department of Physics, 
University of Miami, 
Coral Gables, FL 33124, USA}

\author{Francesca Civano}
\affiliation{Center for Astrophysics | Harvard \& Smithsonian, 
Cambridge, MA 02138, USA}

\author{Fabio Gastaldello}
\affiliation{INAF - IASF Milano, 
via E. Bassini 15, 
I-20133 Milano, Italy}

\author{Roberto Gilli}
\affiliation{INAF – Osservatorio di Astrofisica e Scienza dello Spazio di Bologna, 
Via Gobetti 93/3, 
I-40129 Bologna, Italy}

\author{Fiona Harrison}
\affiliation{Cahill Center for Astronomy and Astrophysics, 
California Institute of Technology, 
Pasadena, CA 91124, USA}

\author{Ann Hornschemeier}
\altaffiliation{Department of Physics and Astronomy, 
Johns Hopkins University, 
3400 N. Charles Street, Baltimore, MD 21218, USA}
\affiliation{NASA Goddard Space Flight Center, 
Code 662, 
Greenbelt, MD 20771, USA}

\author{Ryan Hickox}
\affiliation{Department of Physics and Astronomy, 
Dartmouth College,
6127 Wilder Laboratory, Hanover, NH 03755, USA}

\author{Roman Krivonos}
\affiliation{Space Research Institute, 
Russian Academy of Sciences, 
Profsoyuznaya 84/32, 117997 Moscow, Russia}

\author{Kristin Madsen}
\affiliation{CRESST and X-ray Astrophysics Laboratory, 
NASA Goddard Space Flight Center, 
Greenbelt, MD 20771, USA}

\author{Silvano Molendi}
\affiliation{Istituto di Astrofisica Spaziale e Fisica Cosmica di Milano,
Via Alfonso Corti 12,
20133 Milano, Italy}

\author{Andrew Ptak}
\affiliation{NASA Goddard Space Flight Center, 
Code 662, 
Greenbelt, MD 20771, USA}

\author{Daniel Stern}
\affiliation{Jet Propulsion Laboratory, 
California Institute of Technology, 
Pasedena, CA 91109, USA}

\author{Andreas Zoglauer}
\affiliation{Space Sciences Laboratory, 
University of California, 
Berkeley, CA 94720, USA}



\begin{abstract}

By characterizing the contribution of stray light to large 
datasets from the \nustar\ X-ray observatory collected over 2012--2017,
we report a measurement of the cosmic X-ray background in the 3--20~keV energy range.
These data represent $\sim$20\% sky coverage while avoiding Galactic Ridge X-ray emission 
and 
are less weighted by deep, survey fields than previous measurements with \nustar.
Images in narrow energy bands are stacked in detector space and spatially fit with a model representing the stray light and uniform pattern expected from the cosmic X-ray background and the instrumental background, respectively.
We establish baseline flux values from Earth-occulted data and validate the fitting method on stray light observations of the Crab, which further serve to calibrate the resulting spectra. 
We present independent spectra of the cosmic X-ray background with the FPMA and FPMB detector arrays, which are in excellent agreement with the canonical characterization by \heao\ and are 10\% lower than most subsequent measurements; $F_{\rm{3-20~keV}}^{FPMA} = 2.63 \times 10^{-11}~\rm{erg~s^{-1}~cm^{-2}~deg^{-2}}$ and $F_{\rm{3-20~keV}}^{FPMB} = 2.58 \times 10^{-11}~\rm{erg~s^{-1}~cm^{-2}~deg^{-2}}$.
We discuss these results in light of previous measurements of the cosmic X-ray background and consider the impact of systematic uncertainties on our spectra.

\end{abstract}



\section{Introduction}
\label{sec:intro}

In 1977 NASA launched a low Earth orbiting satellite capable of making nearly continuous large angular measurements of the sky -- \heao. 
Of the four detectors on board, detector A2 was a proportional counter sensitive from 2--60~keV and was specifically designed to measure the CXB, with special attention applied to the separation of the CXB from cosmic and instrumental background signals \citep{Marshall1980-HEAO1-A2,Boldt87}. A more recent reanalysis of this data provides a flux measurement, adjusted to energies from 3--20~keV, of $2.61 \times 10^{-11}$ erg s$^{-1}$ cm$^{-2}$ deg$^{-2}$ \citep{Gruber99}. Various additional measurements have been carried out subsequently, adding to our understanding of the CXB at energies from 0.5--400~keV, all agreeing on the overall shape. The most recent of these include \chandra, \xmm, \swift, \integral, \rxte, \sax, and {\it Insight-HXMT} \citep[e.g.,][]{2006ApJ...645...95H, Cappelluti-2017, Lumb2002-XMM, DeLuca2004-XMM, Ajello2008-swift, Moretti2009-swift/xrt, Churazov2007-integral, Tuler2010-integral, Revnivtsev-2003A&A...411..329R, Frontera2007-BeppoSAX, 2022-Insight}. The overall shape of the CXB in energies from 2--10~keV can be described as a powerlaw with photon index $\Gamma = 1.4-1.52$, and a peak estimated to be between 20--30~keV \citep{CXBshape,Gruber99}. 
While the shape is not in question, the absolute normalization varies $\sim20-30\%$ different measurements. 
This difference is commonly attributed, in part, to the cosmic variance imaging surveys experience due to their smaller solid angle observations \citep{Barcons1992ApJ...396..460B}, and the systematic uncertainties \citep{Churazov2007-integral}. Stray light, improperly modeled, may also contribute to this scatter \citep{Moretti2012-variance}. 

At energies below 10~keV, direct 
measurement and survey field analysis have allowed $\sim80\%$ of the CXB to be resolved at energies $<2$~keV \citep[e.g.,][]{Maccacaro91-EinsteinAGNPOP,Comastri1995-LE/HE,Ueda2003-Chandra+,2007A&A...463...79G,Akylas-le/heAGNpop,Shi2013-LE/AGNpopwithIR,2014ApJ...786..104U,2015MNRAS.451.1892A,2019ApJ...871..240A}. These lower energy observations, when compared to \heao\ A2, have significantly higher reported normalization values, which can not be explained solely due to the cosmic variance \citep[e.g.,][]{Barcons2000-cosmicvariance}. Furthermore, these studies vary from the \heao\ measurements by presenting a flatter slope overall. At energies above 10~keV, before the launch of \nustar, nearly all studies of the CXB were conducted with non-focusing observatories. These studies provided CXB measurements that were consistent in shape, with a``hump", or peak value, at energy values between $20-30$~keV, a few reported higher flux values than \heao\ by $\sim10\%$ \citep[][]{Churazov2007-integral, Ajello2008-swift, Tuler2010-integral}. 

Modern population synthesis models for AGN, whose integrated population densities dominate the CXB, classify AGN based on their hydrogen column density \citep[][]{Setti&Woltjer1989}. AGN are classified as unabsorbed (intrinsic hydrogen column density $N_{H} \leq 10^{22}~{\rm cm}^{-2}$), Compton-thin ($10^{22}~{\rm cm}^{-2} \leq N_{H} \le 10^{24}~{\rm cm}^{-2}$), and Compton-thick ($N_{H} \geq 10^{24}~{\rm cm}^{-2}$). Unobscured and Compton-thin populations fail to recreate the CXB spectral shape or intensity, implying a large population of Compton-thick AGN. To accurately represent these populations, greater certainty in the CXB peak position and normalization is needed.

The Nuclear Spectroscopic Telescope Array (\nustar) provides sensitivity \citep[3--79~keV,][]{2013ApJ...770..103H} that overlaps with the peak of the CXB and covers the energy range (10--20~keV) where the largest disagreement between measurements exists \citep[e.g.,][]{2007A&A...463...79G}.
Early surveys resulted in $\sim$35\% of the 8--24~keV CXB flux being spatially resolved as point sources \citep{2016ApJ...831..185H}.
The remaining unresolved CXB flux can also be measured in these surveys, though, in practice it is difficult to distinguish cosmic photons from more mundane background events of various origin.
However, one source of background is {\it also} due to the CXB: stray light that enters through the aperture stops, completely skirting \nustar's optics.
This stray light background dominates the total background at energies below 15--20~keV and contributes about 10$\times$ more photons per pixel than unresolved CXB sources focused by the optics.
Therefore, the CXB that bypasses the optics and enters through the aperture stops---known as the aperture CXB or aCXB versus the focused CXB or fCXB---provides a more effective way to measure the average CXB spectrum if it can be isolated from other background components.
In addition, this component is independent of the optics and its calibration, providing a more straightforward measurement dependent only on shadowing of the CXB signal by the telescope and the calibration of the detector arrays.

In \cite{2021MNRAS.502.3966K}, a pilot study for this work, the aCXB was isolated from other backgrounds in 
stacked blank-sky survey fields from the
COSMOS \citep{2015ApJ...808..185C}, 
EGS \citep{2007ApJ...660L...1D}, 
ECDFS \citep{2015ApJ...808..184M}, and 
UDS \citep{2018ApJS..235...17M} surveys.
An absolute measurement of the CXB spectrum was made and found to be in agreement with the canonical CXB spectral shape and a total flux $\sim$10\% higher than the \heao\ measurements, consistent with more recent estimates.
Solar influence to the low energy bands up to $\sim7$~keV measured with Earth-occulted data and fit with an XSPEC defined model, {\tt bknpowerlaw}, with $\Gamma_1 \approx 5.0$, $E_{\rm{cutoff}}\approx4.8$~keV, and $\Gamma_2 \approx 0.9$ to correct for this added flux.
However, the source of this exact contribution --- the Sun --- to the on-sky data adds a systematic uncertainty to both the overall normalization and, to a lesser degree, the spectral shape of the CXB due to solar variability and the \nustar\ solar viewing angle.

In this follow-up work, we present a detailed measurement of the CXB intensity and spectrum using the data from 535 observations totaling $\sim$17~Ms per telescope.  
To avoid potential contamination seen in the pilot study, we remove all time periods when the satellite is illuminated by the Sun to get a baseline measurement and apply those parameters to the full time period.
We also calibrate our measured fluxes directly against stray light observations of the Crab \citep{2017ApJ...841...56M} using the same data processing pipeline and method, providing a straightforward way to compare absolute measurements between observatories via cross-calibrations based on the Crab.

The following subsection describes the design of the \nustar\ observatory and its various backgrounds.
In Section~\ref{sec:method}, we describe the process of initial observation selection and exposure time filtering.
Details about how we further select qualifying observations are presented in Section~\ref{sec:selectfit}.
In Section~\ref{sec:Results}, we examine the images of our fully stacked dataset, as well as subsets of the data, to evaluate the fitting procedure and observation selection.
Finally, in Section~\ref{sec:Summary} we discuss our flux measurement and best-fit parameters in light of previous measurements.

\section{\nustar\ Design and Backgrounds}
\label{sec:intro:nustar}



The \nustar\ observatory consists of two co-aligned telescopes each consisting of an optics bench connected with an open mast to a detector bench that contains
a focal plane module (FPM) for each telescope, referenced as FPMA and FPMB \citep[see][for a detailed description]{2013ApJ...770..103H}.
Each detector module incorporates a 2$\times$2 array of monolithic CdZnTe crystals, designated {\tt DET0}, {\tt DET1}, {\tt DET2}, and {\tt DET3} and arranged in a counter-clockwise order starting with {\tt DET0} in the upper right.
Each crystal or detector is electronically pixelated into an array of $32\times32$ ``RAW" pixels \citep{2017SPIE10392E..07G}. 
RAW pixels can each have a unique background response, but in practice each behaves comparably to the others except for those pixels on the edge of a crystal and especially those at the edge of the detector array---the larger surface area of these pixels makes them more likely to trigger on background events. 
Detector-to-detector variations in background arise due to different thicknesses and responsive layers,
which can be characterized in a narrow energy band as a single, uniform instrumental component, $I_c$, for all non-edge RAW pixels in a given detector.
All images in the detector plane are created in DET1 coordinates.

Each location in the focal plane is exposed to a slightly different part of the sky (the solid angle defined by the cone through the aperture stop opening at that position), and a unique fraction of the sky is shadowed by the optics bench.
The changing shadow fraction across the focal planes creates a distinct gradient on each detector array if the source of stray light on the sky is uniform, as is the case with the CXB; we call this component the aCXB or $A_c$ component of the background.
Across a focal plane, the solid angle sampled has an extent approximately 3~degrees in radius.

The optics also focus the diffuse (unresolved) CXB, or fCXB, which is enhanced due to the effective area of mirrors, although the greater solid angle of the aCXB results in $\sim$10$\times$ more photons per pixel being detected from the aCXB relative to the fCXB.
Due to vignetting, the spatial distribution of the fCXB should peak at the location of the optical axis and fall off somewhat with off-axis angle, though scattered light from outside the FOV preferentially falls on the outer parts of the focal plane, effectively filling in the lost emission due to vignetting and creating a flat distribution in the focal plane \citep{2014ApJ...792...48W}.
The fCXB is therefore indistinguishable from the instrumental background in narrow band images and is incorporated in the $I_c$ component.

Lastly, emission from the Sun creates a known soft component that affects the lowest energy end of \nustar's bandpass.
The origin of this component is poorly understood, but it creates a spatial pattern in the focal plane similar to that of the aCXB \citep{2014ApJ...792...48W, 2021MNRAS.502.3966K}.
In \citet{2021MNRAS.502.3966K}, an attempt is made to characterize the solar contribution to the aCXB signal, which appears successful due to the distinct spectral shape of the component.
However, any contribution of that component with a similar spectral shape to the CXB may be confused with the aCXB due to the variability in this particular signal.
This emission can be avoided entirely by selecting observation times when \nustar\ is within the shadow of the Earth, allowing one to characterize the CXB without the added issue of an additional variable in the analysis.


The uniqueness of the shape of the spatial gradient due to the aCXB component allows us to separate out the $A_c$ contribution from the unmodulated $I_c$ component, allowing a relatively straightforward measurement of the CXB with \nustar.

\section{Data Preparation} \label{sec:method}

\nustar\ data was gathered through NASA’s High Energy Astrophysics Science Archive Research Center (HEASARC)\footnote{https://heasarc.gsfc.nasa.gov/W3Browse/nustar/numaster.html}.
We limit observations to those between July 2012 and October 2017; a limited list of 20 Observation IDs (OBSIDs) and cleaned exposure times is given in Table~\ref{tab:obs}.
First, images in the 3-30~keV band for the A and B telescopes were made from the the cleaned event files provided by HEASARC.
These images were searched for sources and circular exclusion regions were made based on the brightness of detected sources, if any were found (details of this procedure are described in Section~\ref{sec:method:srcdet}).
Light curves, binned to 100~s, were then generated
to remove periods of high background; we describe this process in Section~\ref{sec:method:lc}.
Extra background filtering near the SAA was not performed (i.e., the relevant {\tt nupipeline} options set to {\tt SAAMODE=NONE} $\&$ {\tt TENTACLE=NO}), as the light curve filtering step will have already removed high background periods associated with the SAA.
The updated good time interval (GTI) files for FPMA and FPMB used to create new cleaned event files using the {\tt NuSTARDAS} {\tt nupipeline} routine in {\tt HEASoft} version 6.22.1 --- later versions do not introduce any significant changes that would affect our results. Event files were separated into separate files, described further in Section~\ref{sec:method:reproc}, based on the findings presented in Section~\ref{sec:selectfit}.
Images and exposure maps generated from these event files and source masks, which are stacked for many OBSIDs (Section~\ref{sec:method:images}).
The modeling and extraction the CXB signal is described in Section~\ref{sec:method:fitting}.


\begin{deluxetable*}{ccccccccc}
\tablecaption{Limited list of \nustar\ observations used in this work\label{tab:obs}.}
\tablewidth{0pt}
\tablehead{
 & & & \multicolumn2c{On-sky} & \multicolumn2c{Earth-shadowed} & \multicolumn2c{Earth-occulted}\\
 & RA & Dec & A & B & A & B & A & B \\
OBSID & (deg) & (deg) & (ks) & (ks) & (ks) & (ks) & (ks) & (ks)
}
\startdata
  60001113002 & 255.31 & 51.80 & 76.76 & 76.75 & 23.39 & 23.38 & 31.18 & 31.44  \\
  60101073002 & 181.67 &-31.94 & 27.69 & 27.45 &  2.22 &  2.22 & 11.84 & 11.81  \\
  60001120002 & 121.12 & 65.00 & 27.50 & 27.38 &  8.47 &  8.47 &  0.00 &  0.00  \\
  60101065002 & 183.80 &-14.50 & 26.48 & 26.46 &  2.97 &  2.96 & 16.14 & 16.12  \\
  60361023002 & 233.87 & 73.46 & 26.31 & 26.51 & 10.14 & 10.40 &  0.00 &  0.00  \\
  60101082002 & 196.27 &  0.91 & 26.14 & 26.25 &  6.13 &  6.13 & 18.02 & 18.09  \\
  50101005002 & 175.44 & 32.28 & 26.44 & 25.91 &  3.65 &  3.64 & 15.32 & 15.31  \\
  60101064002 & 182.41 & -5.03 & 25.09 & 25.07 &  4.97 &  5.08 & 11.64 & 11.82  \\
  60110003003 & 189.13 & 62.20 & 59.72 & 59.56 & 15.77 & 15.92 & 17.30 & 17.45  \\
  60110003004 & 189.16 & 62.20 & 60.18 & 59.97 & 13.75 & 13.71 &  8.23 &  8.20  \\
  60001107002 & 156.62 & 25.73 & 59.10 & 59.05 & 22.93 & 22.92 & 36.81 & 36.52  \\
  60110002004 & 189.23 & 62.23 & 57.20 & 57.40 & 13.02 & 13.08 &  8.14 &  8.10  \\
  60110001003 & 189.28 & 62.28 & 55.04 & 55.00 & 17.73 & 17.68 &  9.55 &  9.43  \\
  60101004002 & 328.74 & -9.40 & 53.89 & 53.58 & 13.51 & 13.46 & 39.26 & 39.43  \\
  60110001005 & 189.31 & 62.27 & 53.77 & 53.85 & 10.41 & 10.38 &  8.50 &  8.47  \\
  60001136002 & 190.87 & -2.54 & 50.75 & 50.97 &  9.07 &  9.06 & 30.58 & 30.45  \\
  60366002002 & 352.63 & -0.65 & 50.84 & 50.46 & 17.94 & 17.86 & 28.19 & 28.33  \\
  60110003005 & 189.16 & 62.19 & 49.61 & 49.57 & 21.10 & 21.06 &  9.29 &  9.37  \\
  60110001007 & 189.31 & 62.27 & 48.27 & 48.71 & 20.87 & 21.01 & 10.79 & 10.98  \\
  60110002005 & 189.24 & 62.23 & 48.60 & 48.68 & 20.90 & 20.84 & 10.18 & 10.20  \\
\enddata
\tablecomments{On-sky columns indicate the cleaned exposure time for periods when the satellite is on target conducting science-mode observations. Earth-shadowed being a subset of that period where the satellite made observations while in Earth's shadow. Earth-occulted columns list the times during those observation the sky is blocked by the Earth. Total number of observations for On-sky is 535, of those, 329 had sources within the field of view while 206 had no source detectable. The total number of observations for Earth-occulted is 959.}
\end{deluxetable*}

\subsection{Source Detection}
\label{sec:method:srcdet}

\begin{figure}[h]
    \centering
    \includegraphics[scale=0.35]{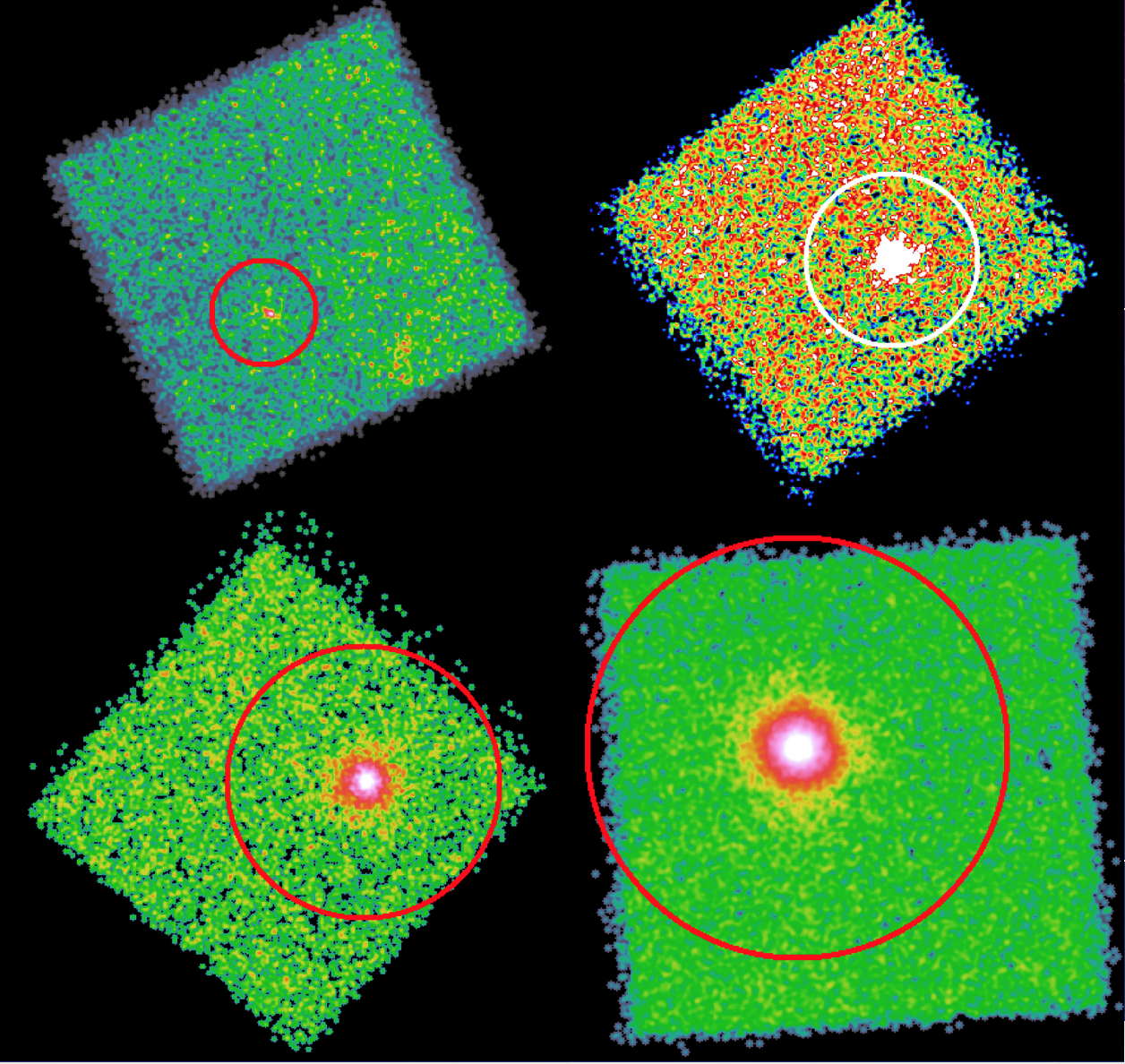}
    \caption{Combined images from telescope A and B in the energy band of 3-30~keV displaying examples of exclusion regions based on the source detection program described in Section \ref{sec:method:srcdet}. Left to right, top to bottom displays the smallest to largest sources found that qualified for this study with an observation ID and radii given, in pixels, 60101101002 at 47.85 pixels, 60001134002 at 56.65 pixels, as 90201019002 at 108.35 pixels, and 60102051006 at 155.65 pixels respectively.}
    \label{fig:srcex}
\end{figure}

Projected images on the sky---in the {\tt SKY} coordinate system---from telescope A and B for each observation over the range $3~{\rm keV} < E < 30$~keV are first summed to create a single image for the observation. That image is then convolved with a Gaussian kernel ($\sigma = 6$~pixels). Using the exposure-weighted background level from the \nustar\ background study \citep{2014ApJ...792...48W}, 
we identified any pixels 5$\times$ above the local background level.
In order of descending pixel value above the threshold, a circular exclusion region is centered on the pixel
Circular exclusion regions, centered on the point source, was applied to each central pixel that exceeded the background threshold.
The radius of the exclusion region was set by where the point spread function (PSF), its peak scaled to the pixel value, falls to 3\% of the background level.
The assumed PSF is the on-axis, lowest energy version stored in the {\tt CALDB}.
This creates an exclusion region many times the size of the visible source in a single observation (Fig.~\ref{fig:srcex}), but avoids the wings of \nustar's PSF which become detectable in stacked observations if a smaller exclusion region is used.
Most \nustar\ observations target point sources that lie in the same area of the detector planes.
Our approach allows us 
to consider each observation to be ``source free," with the only remaining cosmic source of photons being the CXB, both from focused, undetected sources across the FOV and from unfocused, stray light through the aperture stops.

To verify the reliability of our source detection methodology, we cross-checked our list of sources with the most recent study done by the \nustar\ serendipitous survey team \citep[][submitted]{Klindt22}. 
All observations where a source was detected and appear in the 80-month survey showed total agreement.
This agreement gives us confidence that we have removed the majority of detectable individual sources and that any faint ones we may have missed will contribute negligible source counts relative to our systematic uncertainties, especially since small localized excesses do not correlate with the shape of the full FOV gradient used to extract the CXB signal.
There were discrepancies in our comparison that should be noted. The serendipitous survey excluded all targeted blank field surveys (e.g., COSMOS, UDS, EGC, and ECDFS) where in a select few observations, we detected sources. A few observations we originally included were excluded in the serendipitous survey due to excess background contamination such as stray light and ghost rays. These observations were excluded from our final survey catalog and do not contribute to our measurement. 


\subsection{Light Curve Filtering}
\label{sec:method:lc}


Background light curves, extracted from the entire FOVs except for those regions with sources as described in Section~\ref{sec:method:srcdet}, 
would ideally have no temporal variation.
However, SAA passages temporarily increase the instrumental background, largely at high energies, and enhanced solar activity can increase the background at lower energies.
These variations occur on generally short (few minute) timescales compared to typical observations, which span a bit under a day to several days in total time.
As such, these ``flaring" periods can be readily identified as deviations from the mean background rate and removed via sigma-clipping.
However, some light curves, even those without flaring events, exhibit an overall sinusoidal variation with a period of about a day.
The amplitude of the variation is not constant across observations, with some light curves showing essentially no variation.
While the origin of this variation has not been determined, it needs to be accounted for so that the dispersion from the mean of the light curve is not overestimated, which would cause some low level flaring to be missed.

\begin{figure*}[ht]
    \centering
    \includegraphics[width=\textwidth]{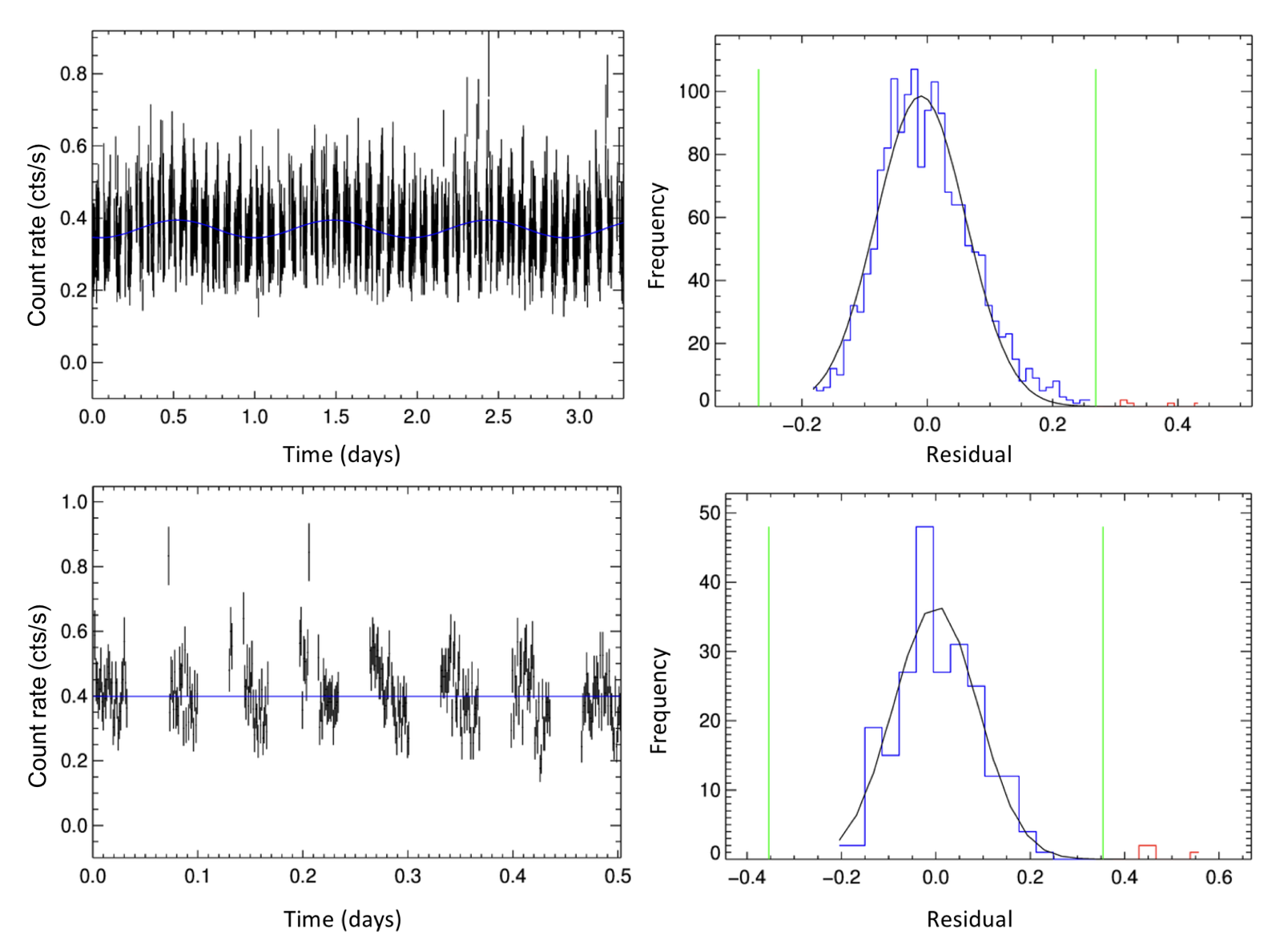}
    \caption{Left panels: Light curves binned in 100 second intervals, shown with the best model fit (solid blue line). Right panels: histograms showing the distribution of residual bin values around the best fit line, fitted by a Gaussian curve.  The 3.5$\sigma$ filtering thresholds are shown as green lines on either side of the distribution; red bins indicate those time intervals removed from the GTI file.
    The top and bottom panels show OBSID 50002019004 (FPMA) and OBSID 60001068002 (FPMB), respectively.}
    \label{fig:lc}
\end{figure*}

Established protocols for handling the exclusion of flaring events were used to minimize systematic issues in the initial processing.
After the source detection for an observation, the event file for selected energies from 50-100~keV were binned into 100 second selections for the interval of the observation as seen in the left panels of Figure~\ref{fig:lc}.
In this bandpass, the contribution of counts from most sources is negligible compared to the background.
The 50-100~keV energy band corresponds to moments of high flux rates caused by high energy particles interacting with the spacecraft causing many activation lines. Gaps between event periods are due to occulations of the sky by the Earth (these light curves exclude Earth-occulted data).
Three different functions (linear with zero slope, linear with non-zero slope, and sinusoidal) were fit to the light curve in a least-squares minimization program as part of the {\tt Lmfit} {\tt python} package\footnote{https://lmfit.github.io/lmfit-py} and the most appropriate function is selected as the one with the lowest $\chi^2$ value.

Assuming the functional form is correct and that there are sufficient counts per bin (ensured by the bin size, typically yielding $\sim$40 counts/bin, Fig.~\ref{fig:lc}),
residuals around the best-fit model will follow a Gaussian distribution.
We fit the distribution of residuals with a Gaussian function and set a threshold value of 3.5$\sigma$ beyond which that time interval is excluded from the light curve and GTI file.
This procedure is repeated on the same light curve in case the original fit was biased by solar flares.
Finally, the exclusion region due to a detected source, if one was found, is used to filter events to exclude source counts in light curves in the $3~{\rm keV} < E < 7$~keV energy band.
The above fitting and sigma-clipping procedure is then applied to the lower energy light curves in order
to remove any flaring associated with the Sun.
Flagged time periods are again removed from the GTI file.

Data from FPMA and FPMB are treated as separate observations, and the light curves and GTI files are processed independently.
Although high background periods should arise in the light curves of both focal planes simultaneously, the appearance of flares are not quite identical.
Treating the datasets independently maximizes the accepted exposure time and provides two somewhat uncorrelated measurements of the CXB.

\subsection{Event File Cleaning and Separation}
\label{sec:method:reproc}

\begin{figure}[t]
    \centering
    \includegraphics[width=\linewidth]{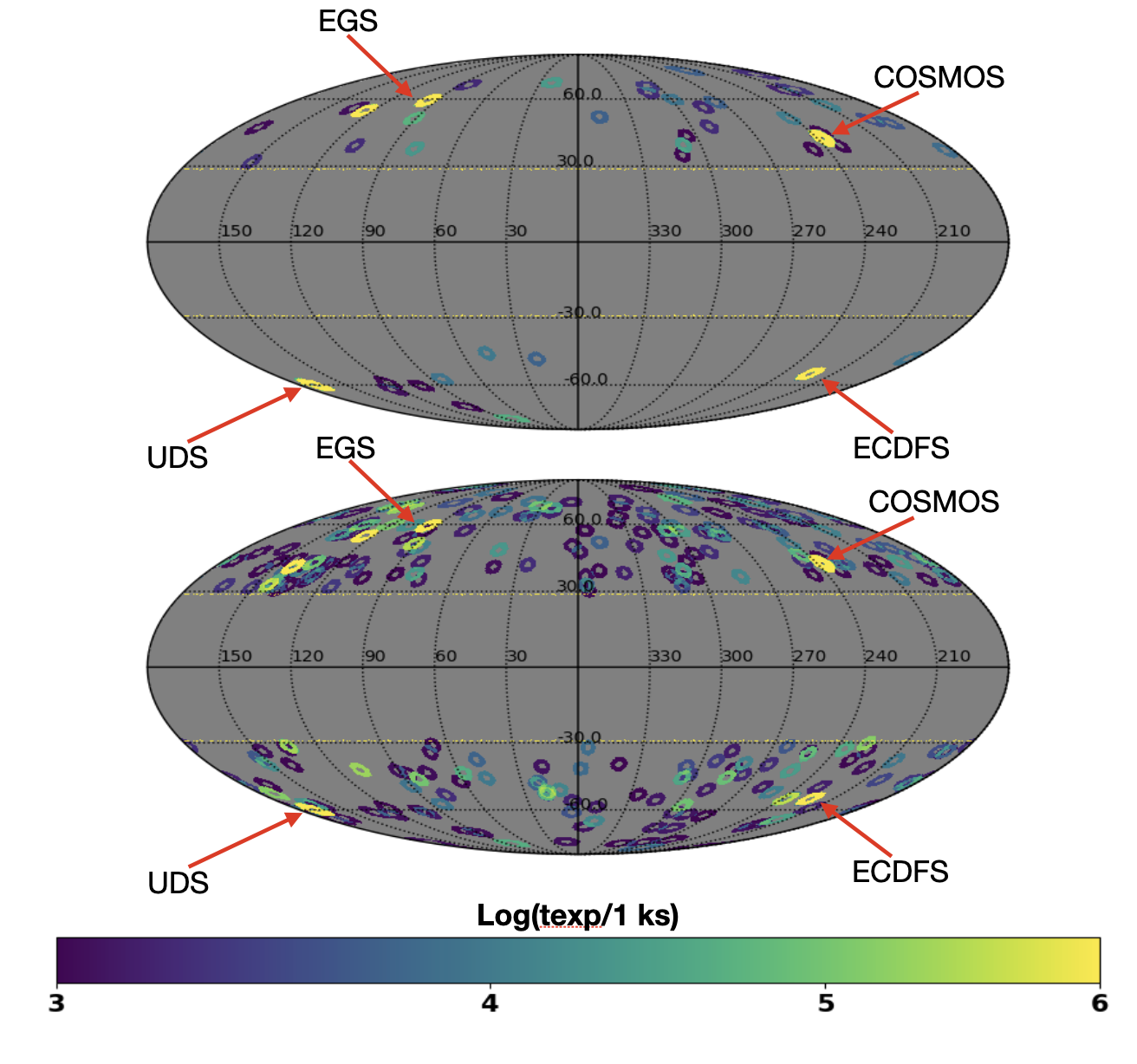}
    \caption{Mollweide projection in Galactic coordinates displaying the qualifying observations and overall total exposure time for the Blank Sky list (top) and the Full Sky list (bottom). The toroidal shape of the points reflect the summed area of the sky visible to both telescopes A and B by stray light.}
    \label{fig:mollweide}
\end{figure}

Using the light curve-filtered GTIs, new cleaned event files are generated from the original unfiltered event files with {\tt nupipeline}.
This reprocessing produces two relevant event files per telescope: the on-sky or science mode events, when \nustar\ has a clear line of sight to the target of interest, and the Earth-occulted events, when the \nustar\ boresight is pointed at the Earth.
These cleaned event files are
further separated into time periods when the spacecraft is either illuminated by the Sun or in Earth's shadow, filtered with the {\tt nustardas} routine {\tt nuscreen} via the {\tt SUNSHINE} flag.
These event files are used to generate images from each observation and focal plane that are then stacked after OBSID selection criteria (Section~\ref{sec:selectfit}) are applied.
We separately consider several collections of OBSIDs: ``Blank Sky," observations where no sources were detected, four surveys \citep[EGS, COSMOS, UDS, and ECDFS:][respectively]{2007ApJ...660L...1D,2015ApJ...808..185C,2015ApJ...808..184M,2018ApJS..235...17M}, which are largely, but not entirely contained within the ``Blank Sky" set, and ``Full Set," the total collection of OBSIDs meeting our selection criteria, including the preceeding collections.
The on-sky pointings of the Full Set OBSID list, described in Section~\ref{sec:selectfit}, is presented in Figure~\ref{fig:mollweide}. The approximate solid angle of the aCXB stray light is represented with a $\sim$1\arcdeg---3\arcdeg\ annulus, color-coded by exposure time.

\subsection{Image Stacking and Exposure Maps}
\label{sec:method:images}



\begin{figure}[t]
    \centering
    \includegraphics[width=\linewidth]{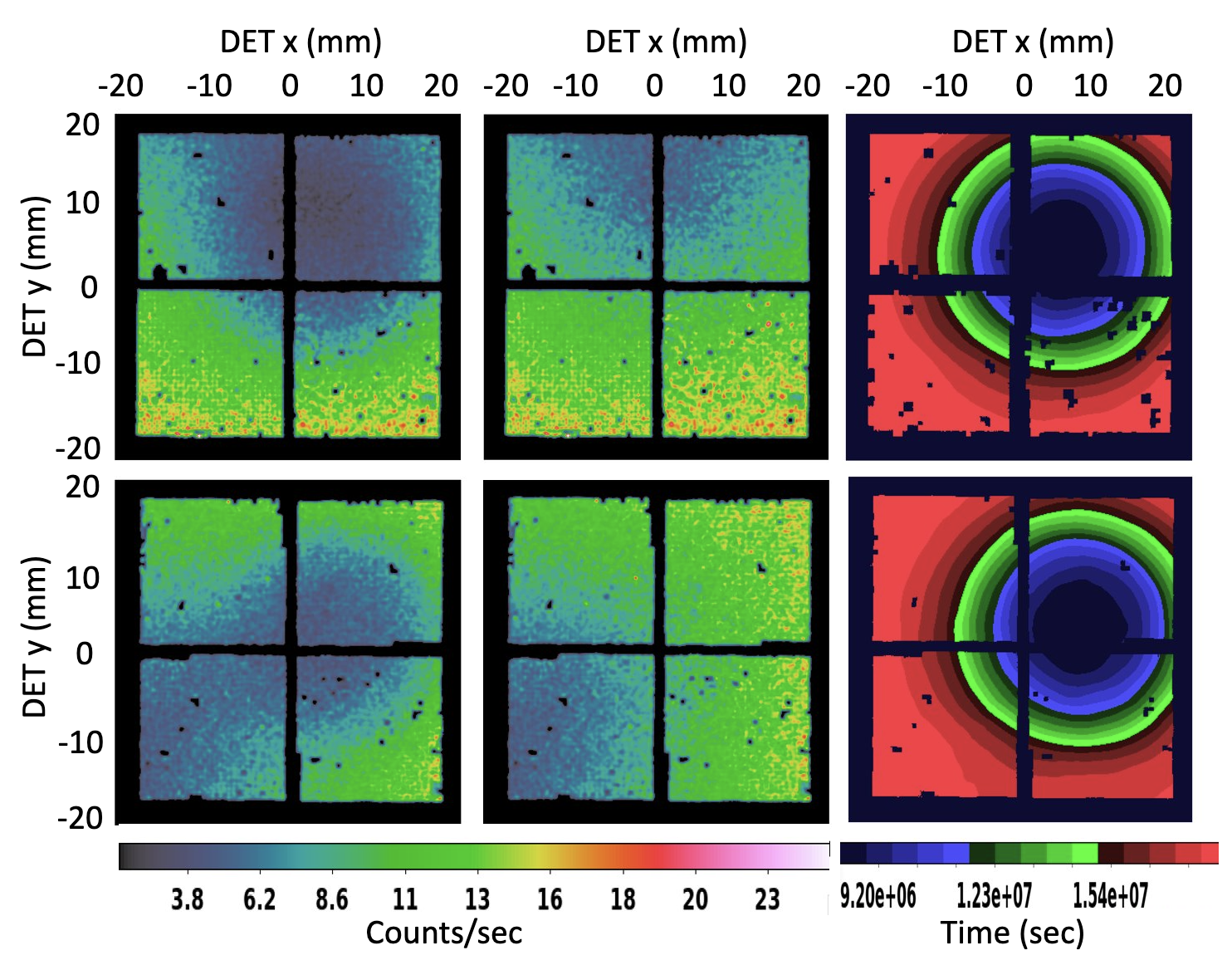}
    \caption{
    Left and Middle Columns: Stacked science data in the 3--12~keV energy band with the Full Set OBSID list for FPMA (top) and FPMB (bottom) for a total exposure time of $\sim18.53$~Ms and $\sim18.46$~Ms respectively.
    The left images show raw stacked counts, and the middle images are exposure-corrected versions with the maximum exposure time normalized to 1.
    Right Column: Stacked exposure maps for all observations used with this source list. 
    }
    \label{fig:Fullsky_datamodel}
\end{figure}

Images in a particular energy band are created from cleaned event files in the DET1 coordinate system, and all model fits are also performed with the data projected in these coordinates.
Exposure maps are in principle straightforward; because vignetting (or in this case shadowing of the CXB by the optics bench) is encoded in the aCXB spatial model, all DET1 pixels should share the exposure time of the observation.
However, some RAW pixels are faulty and turned off and other ``hot" RAW pixels experience occasional, extra triggering that causes them to record more events than their surrounding pixels.
In stacks of large numbers of OBSIDs (simply summing the images of all the observations together), RAW pixels not identified in bad pixel lists became obvious.
The corresponding DET1 pixels of those RAW pixels were identified and added to the list of known bad pixels.
In addition, all RAW pixels on the edges of detectors are included in the list of bad pixels, for the reasons discussed in Section~\ref{sec:intro:nustar}.
The bad pixel list is then used to create a mask that zeros out these pixels in the exposure map.
For observations without identified sources, 
the data and exposure images are ready to be summed with those from other observations.
Figure~\ref{fig:Fullsky_datamodel} shows an example of stacked and exposure-corrected FPMA and FPMB data images for the Full Set OBSID list, on sky, when \nustar\ is in Earth's shadow.

For observations with identified sources and corresponding exclusion regions (see Section~\ref{sec:method:srcdet} for details), only events falling outside the exclusion region should be included in the data image, and the exposure map should account for the fraction of time pixels lie outside the region.
Because the exclusion region is defined in sky coordinates, we trace the exclusion region from the sky to the focal plane DET1 coordinates using the same data that are used to project the source photons onto the sky.
Relative motions between the optics and detector planes cause the position of the optical axis---and thus source positions---to shift over the course of an observation.
These shifts are recorded by the laser metrology system in typically quarter-second intervals; the exclusion region is recentered in DET1 coordinates for each of these intervals, and all pixels that fall within the region have their exposure time reduced by the time interval.
Events are similarly removed from the data image, although the procedure is more straightforward since they can be directly filtered on their DET1 coordinates.
The bad pixel mask is then applied to the exposure map, and both the data and exposure images are ready to be stacked with other observations through simple addition.

The resulting exposure maps are energy-independent; energy dependent effects, such as absorption by the beryllium window or the dead layer on the top surface of the crystals, are accounted for during spectral fitting in the response files.
For each narrow band, the model image is multiplied by the exposure map to produce predicted counts per pixel that can be directly compared to the data stack.


\subsection{Image Fitting and Construction of Spectra}
\label{sec:method:fitting}

\subsubsection{Aperture Model}
\label{sec:method:fitting:ap}

Over the large solid angles that \nustar\ is sensitive to, the CXB has a uniform brightness across the entirety of the sky.
This isotropic signal simplifies the modeling needed and implies that each pixel is only dependent on 
the solid angle of the sky visible to it.
This angle is defined by the relative position of the aperture stop, which allows each pixel to possibly view $\Omega \sim 12~{\rm deg}^2$ based on the geometric size and distance to the stop. 
Depending on the location of a particular pixel, some fraction of that solid angle is obscured by the optics bench.
The spatial model for the aCXB, $A_c(x,y)$, as a function of DET1 pixel location $(x,y)$, is provided by the {\tt nuskybgd} code distribution \citep{2014ApJ...792...48W} and successfully used for this purpose in \citet{2021MNRAS.502.3966K}.
The solid angle at each pixel is computed using a simple ray trace through the geometry of the telescope, which assumes the detector array is centered directly below the 58~mm diameter outermost aperture stop, placed 833.2~mm from the focal plane, which is 10.15~m distant from the optics bench.
The relative positions of the detector planes, aperture stops, and optics bench are all consistent with the stray light patterns of known, bright sources \citep{2017ApJ...841...56M, 2021ApJ...909...30G}.

\begin{figure}[t]
    \centering
    \includegraphics[width=\linewidth]{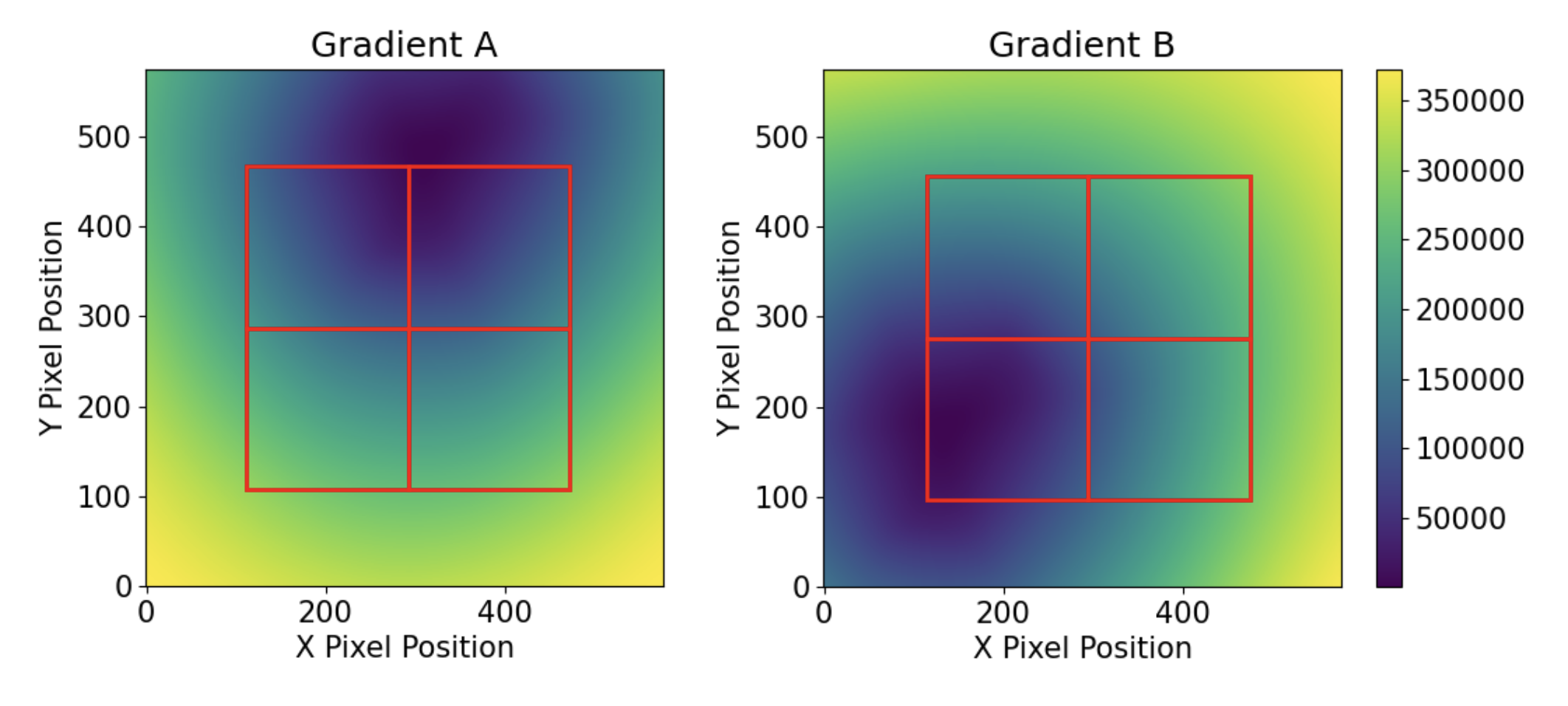}
    \caption{The aCXB gradient model incorporated in {\tt nuskybgd} for FPMA (left) and FPMB (right). The red squares represent the locations of each detector including some padding inside the entire 360$\times$360 pixel focal plane defined by DET1 coordinates (left column of Fig.~\ref{fig:Fullsky_datamodel}). Each image ranges in pixel value from $\sim440$~mm$^2$ (dark blue) to $\sim370,000$~mm$^2$ (yellow).
    These areas scale with the solid angle visible to each pixel and correspond to the area projected onto the optics bench plane.}
    \label{fig:grads}
\end{figure}

While the size of the aperture stop, outline of the optics bench, and the distance of both from the detector plane are unlikely to be different after launch from that designed and manufactured, a tiny misalignment in the positions of the detectors relative to the aperture stops is perhaps most possible.
To evaluate this possibility, the positions of stray light patterns created by known sources, taken from {\tt StrayCats}\footnote{https://nustarstraycats.github.io/straycats/tables/straycats\_table} \citep{2021ApJ...909...30G}, were compared to the predicted positions of the pattern using the same code that generated the aCXB models.
We use the stray light patterns with {\tt STRAYIDs~StrayCats\_I} 53, 154, 260, 262, 264, 269, 275, 277, 279, 293, 404, 652, and 675 for FPMA and {\tt STRAYIDs~StrayCats\_I} 2, 261, 263, 265, 270, 276, 278, 280, 302, 306, 334, 405, and 676 for FPMB. 
In {\tt ds9}, circular regions with the size of the aperture stop are aligned with the observed and predicted stray light patterns by eye, which yields sub-pixel precision, and the difference between the centers of the two circles is recorded.
Differential thermal expansion of the mast---which varies from one observation to another---causes a small angular shift between the relative orientation of the optics and focal plane modules, corresponding to shifts between the predicted and observed positions of the stray light: around $\pm$5~pixels in one direction.
The sense and amount of this shift depends on the orientation of the spacecraft relative to the Sun; assuming these observations are representative of the spread in orientations for our datasets, and accounting for observations of nearby fields in the above list that have similar shifts to each other, we find the aCXB pattern should be shifted by $(\Delta x,\Delta y)$ of $(8, -5)$ for FPMA and $(10, -16)$ for FPMB in the DET1 coordinate plane.
We apply these shifts to our aCXB model component $A_c (x,y)$ during all fits.
In principle, these shifts could be left as free parameters in the fit, but the strong degeneracy between shifts along lines of symmetry and the normalization of the component makes this option impractical.
The locations of the detectors, including the above shifts, relative to the {\tt nuskybgd} aCXB spatial gradient models are shown in Figure~\ref{fig:grads}.
The underlying gradient represents the solid angle through the aperture stop on the sky visible to a pixel at that location.
The {\tt nuskybgd} images are computed at the optics bench plane and have units of mm$^2$; given the focal length of \nustar, 1~mm$^2$ corresponds to a solid angle of $3.1865\times10^{-5}$~deg$^2$.


\subsubsection{Total Model}
\label{sec:method:fitting:model}

To extract the CXB flux from an image, we construct a model that includes the aCXB gradient, $A_c(x,y)$, and individual flat components for each detector to account for the instrumental and fCXB background, $I_{c,i}$, where $i$ corresponds to the four detectors DET0, DET1, DET2, and DET3:
\begin{equation}
\begin{aligned}
    M(E,x,y) = {} & a_0(E)I_{c,0}(x,y) + a_1(E)I_{c,1}(x,y) + \\ 
    & a_2(E)I_{c,2}(x,y) + a_3(E)I_{c,3}(x,y) + \\
    & N_{a}(E)A_{c}(x,y)\, ,
\end{aligned}
\label{eq:imfit}
\end{equation}
where $M(E,x,y)$ is our overall model count rate as a function of pixel position $x,y$ for energy band $E$, $a_0(E)$ through $a_3(E)$ are the respective count rates for the instrumental component of each detector, $I_{c,i}(x,y)$ are masks identifying which pixels correspond to detector $i$, 
$N_a(E)$ is the CXB count rate per deg$^{2}$, and $A_{c}(x,y)$ is the spatial gradient of the aCXB in deg$^{2}$ across the focal plane.
In order to fit the model parameters to the data with the Cash statistic likelihood
\citep{1979ApJ...228..939C,2007A&A...472...21L,2009ApJ...693..822H,2017A&A...605A..51K},
pixels are combined to form bins with at least 2 counts per bin
to avoid improper arguments for the logarithms:
\begin{equation}
    C = 2\sum_{i=1}^{N} \Big\lbrack (t M_{i}) - S_{i} + S_{i}(\ln(S_{i})-\ln(t M_{i})) \Big\rbrack\, ,
    \label{eq:cash}
\end{equation}
where $i$ is the bin index, $t$ is the summed exposure time for the pixels in the bin from the exposure map, $S_{i}$ is the observed counts from the summed pixels in the data image, and $M_{i}$ is the summed model rate from all pixels in the bin following Equation~\ref{eq:imfit}.
The best-fit parameters of the model are found by minimizing $C$ using a
Nelder-Mead optimization function. This function consistently returned lower $C$ values when compared to other methods and was found to be the most self-consistent optimization method.

\subsubsection{Spectra and Verification}
\label{sec:method:fitting:spec}

\begin{figure}[t]
    \centering
    \includegraphics[width=\linewidth]{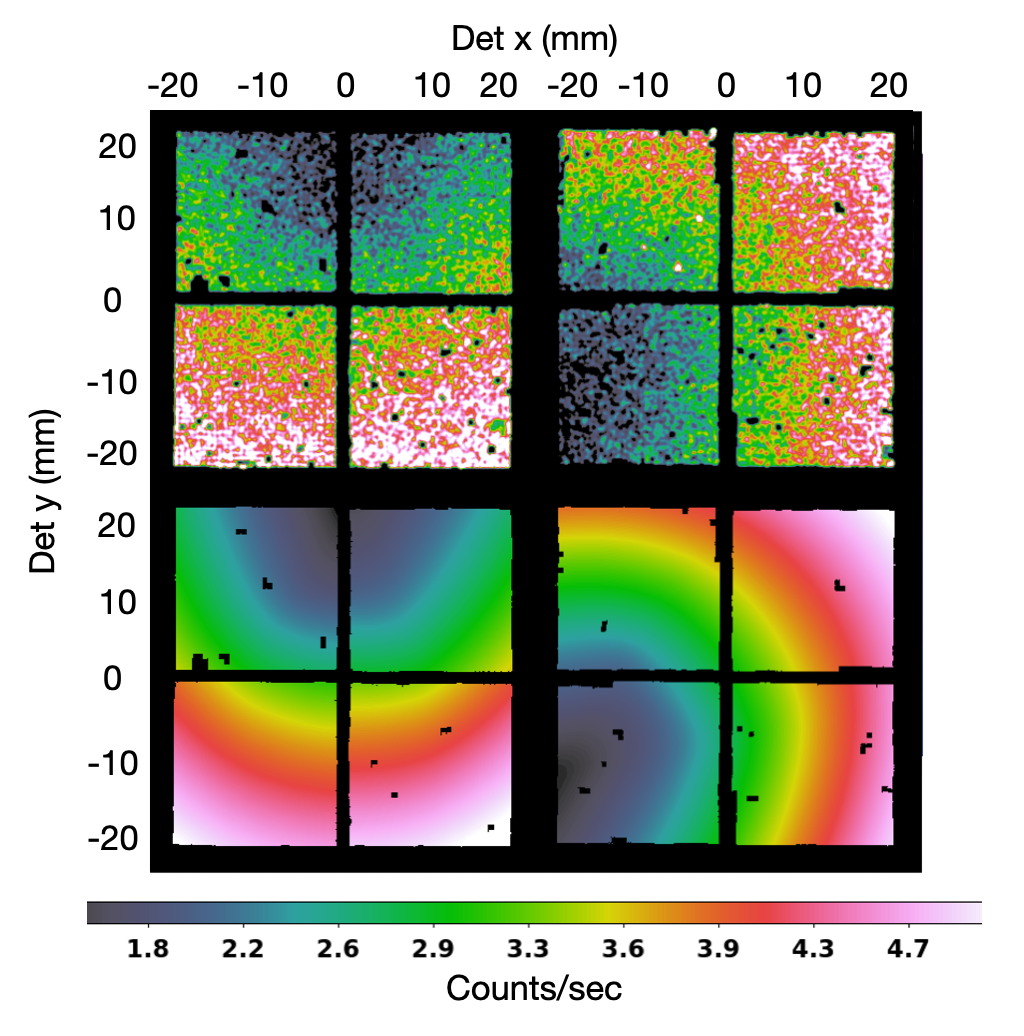}
    \caption{Top: Stacked data for the Blank 
    Sky list in the energy band 3--8~keV with a total exposure of $\sim7$~Ms for each telescope. Bottom: Model images created from the best-fit parameters of Equation~\ref{eq:imfit} multiplied by the exposure maps.
    Slight offsets in the gradient pattern are due to differing instrumental $a_i$ contributions.
    FPMA and FPMB are shown in the left and right columns, respectively.}
    \label{fig:Blanksky_datamodel}
\end{figure}

Energy bands are spaced in increasing bin widths starting at a minimum of 3.00 -- 3.20~keV at the lowest energies to 38.24 -- 41.24~keV at the highest energies. Energy bin widths much smaller than this did not return a consistent signal, while larger binning gave expected similar normalizations of the aCXB signal.
Figure~\ref{fig:Blanksky_datamodel} shows the Blank Sky data stack (top row) and model (bottom row) images in the 3--12~keV energy band for both FPMA and FPMB (left and right columns, respectively).  The exposure maps in this case are uniform except for detector gaps and masked pixels, which are visible in the model images.  Uncertainty ranges are then calculated by stepping $N_{a}$ away from its best fit value and refitting the $a_i$ detector values through our optimization function until the change in the C-statistic, $\Delta C$, increases to the square of the desired sigma (generally 1$\sigma$ for spectra). As reported in \citet{C-stat-humphrey-2009}, using this form of the C-statistic --- which is the primary form used in {\tt XSPEC} \citep[][]{1996ASPC..101...17A} --- allows us to calculate the errors.
The CXB spectrum for each energy band were measured by repeating this process.

To verify our fitting procedure and provide a useful cross-calibration to absolute flux, we reproduced the efforts of \citet{2017ApJ...841...56M} to measure the stray light fluxes produced by the Crab nebula in several \nustar\ calibration observations. The only variation to our method was the introduction of a single component that represents the Crab stray light:
\begin{equation}
\begin{aligned}
    M(E,x,y) ={} & a_0(E)I_{c,0}(x,y) + a_1(E)I_{c,1}(x,y) + \\ 
    & a_2(E)I_{c,2}(x,y) + a_3(E)I_{c,3}(x,y) + \\
    & N_{a}(E)A_{c}(x,y) + c_c(E)C(x,y)\, ,
\end{aligned}
\end{equation}

\begin{figure*}[ht]
    \centering
    \includegraphics[width=\linewidth]{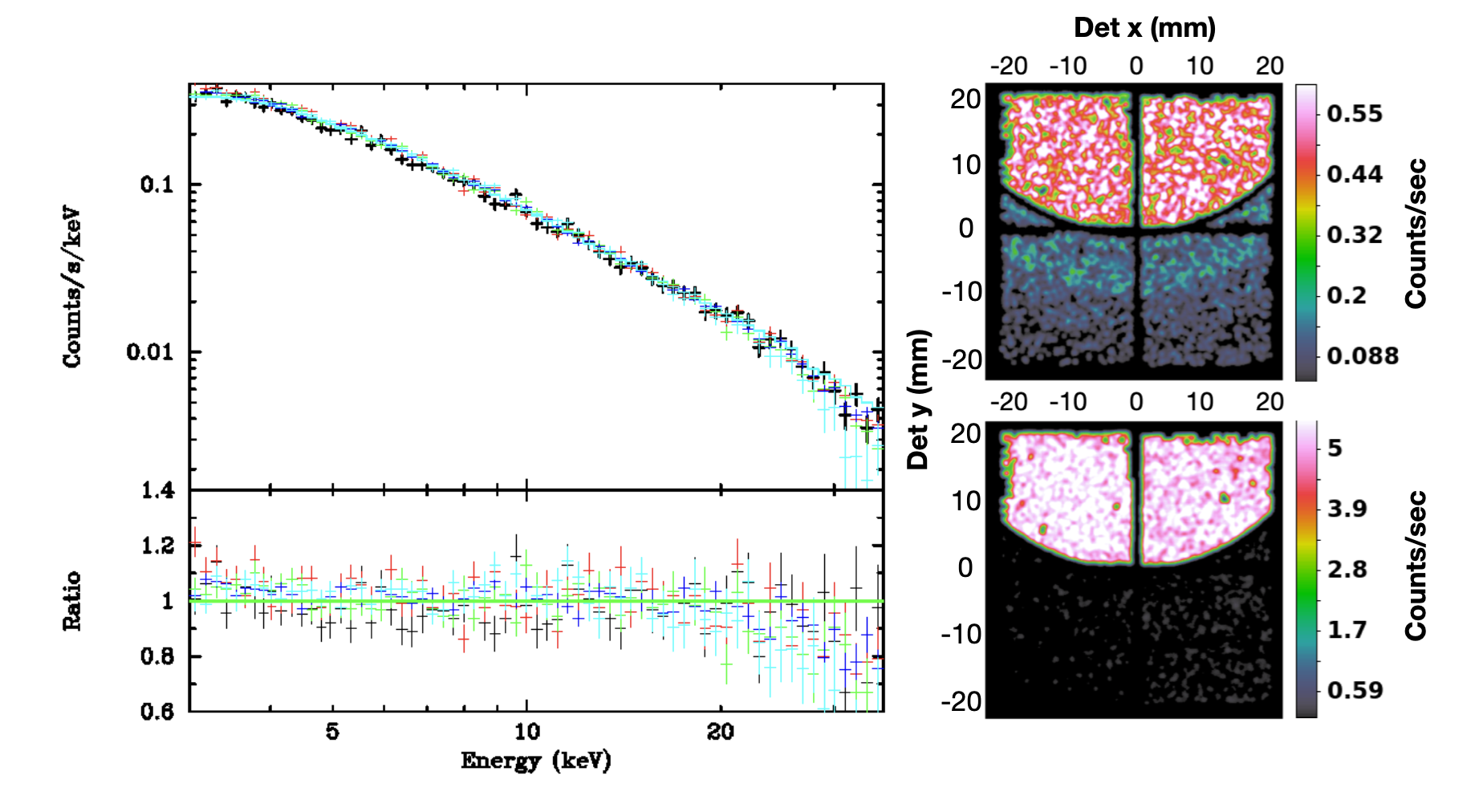}
    \caption{Left: Spectra extracted from stray light observations of the Crab fit to a power law model.  The data points are derived from fits to narrow-band images performed in the same way as in the CXB analysis. Right: Detector images for OBSID 10110005001 of FPMB in the 20--40~keV (top) and 3--20~keV (below) energy ranges. Above 20~keV the aperture stops start to become transparent, as evidenced by the additional half-circle stray light patterns seen in the top image. More information can be found in \citet{Madsen-straylight}}
    \label{fig:crab}
\end{figure*}

where $c_c(E)$ is the Crab count rate and $C(x,y)$ is the mask indicating which pixels contain stray light from the Crab. Our Crab measurement utilized CALDB version 20191219 --- vs the 20170727 used for the observations. The difference between these two versions account for a long term gain calibration, but do not effect on our measurement in any way. Our optimization was performed over four different observations with five different images: 10110001002 FPMA and FPMB, 10110003002 FPMB, 10110004002 FPMA, and 10110005001 FPMB. Compared to similar measurements of \citet{2017ApJ...841...56M}, we find a 1.5\% ($3.316\pm0.010$~x~$10^{-8}$ erg s$^{-1}$ cm$^{-2}$) lower flux when fitting energies up to 22~keV and a 3.6\% ($3.2505\pm0.011$~x~$10^{-8}$ erg s$^{-1}$ cm$^{-2}$) lower flux when fit up to 40~keV. The measured photon index was consistent within $1\%$ in both energy ranges with values of $\Gamma_{3-22} = 2.076\pm0.007$ and $\Gamma_{3-40} = 2.103\pm0.006$. 
The spectra and fit are shown in Figure~\ref{fig:crab}. Additional stray light can be seen in energies above 20~keV --- the half-circular shape below the strong stray light pattern in the upper right image of the figure. This extra light biases spectral fits above 20~keV as the image-fitting model does not include it, leading to an overestimate of the detector background and thus underestimate of the Crab flux.


\section{Data Selection} 
\label{sec:selectfit}

Observations were initially selected to avoid those containing extended sources (e.g., targeted galaxy cluster observations, which have OBSIDs beginning with 7), stray light contaminated sources, and bright sources where the source exclusion region determined in Section~\ref{sec:method:srcdet} has a radius of larger than 180 pixels and excludes more than 2/3 of the image in the {\tt DET1} coordinate frame.
In addition, observations near the plane of
the Milky Way, with Galactic latitude $|b| < 10^{\circ}$, were removed in an effort to avoid diffuse signal from our own Galaxy, i.e., from the Galactic Ridge X-ray Emission \citep[GRXE, e.g.,][]{Valinia1997GRXE,Revnivtsev2006,KrivonosGRXE}.

\begin{figure}[ht]
    \centering
    \begin{subfigure}
        \centering
        \includegraphics[width=.4\textwidth]{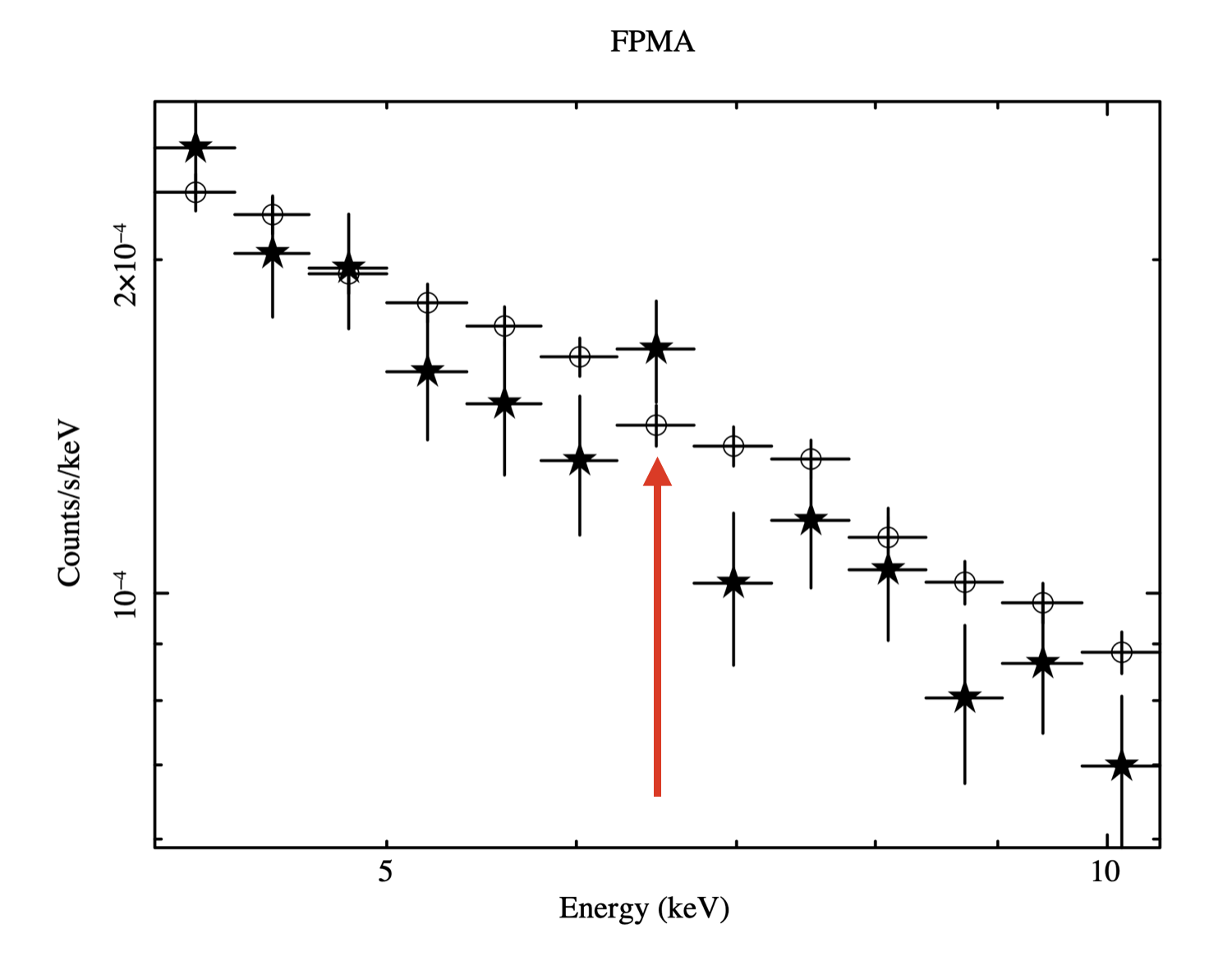}
        \label{fig:A64}
    \end{subfigure}
    \begin{subfigure}
        \centering
        \includegraphics[width=.4\textwidth]{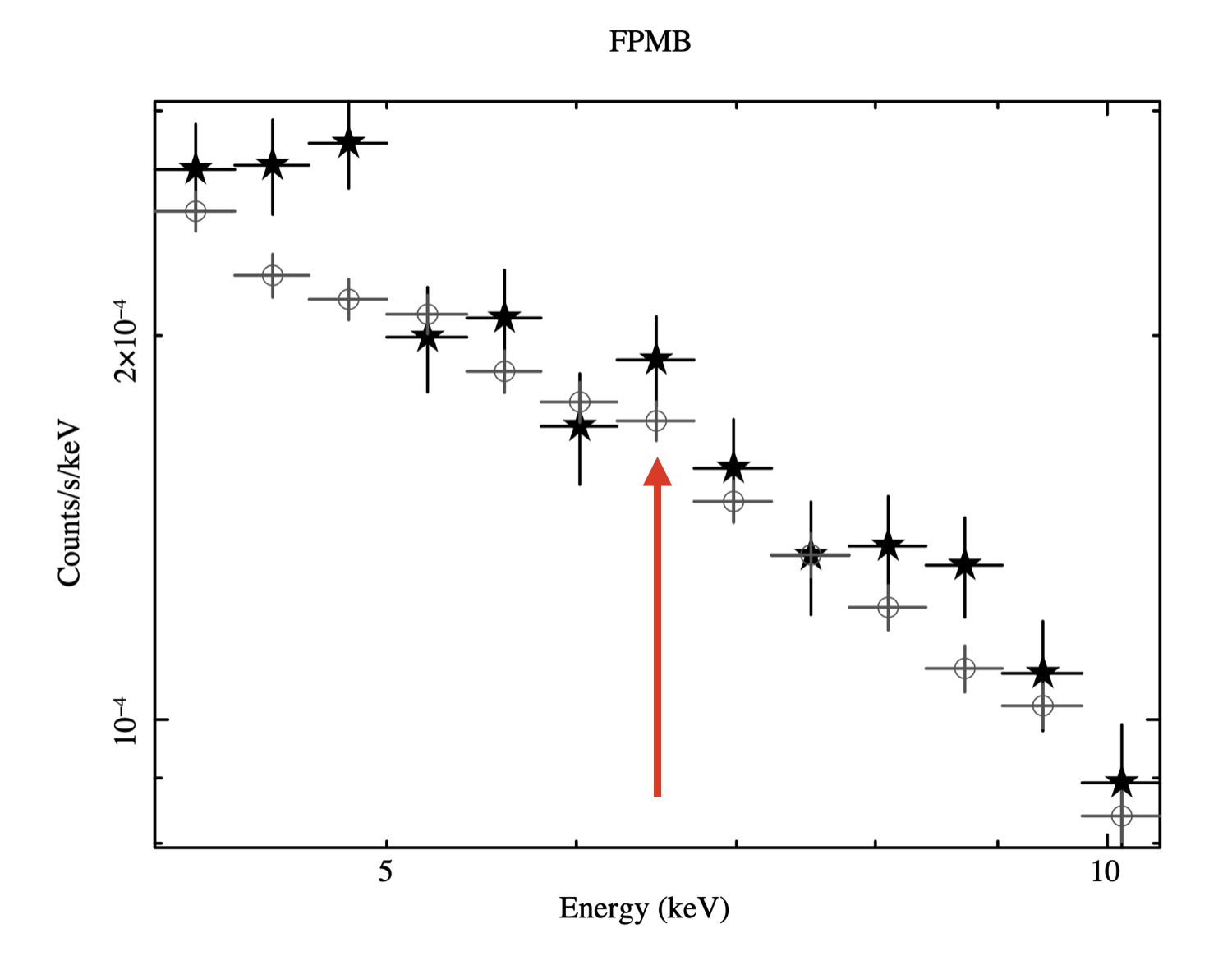}
        \label{fig:B64}
    \end{subfigure}
    \caption{FPMA (top) and FPMB (bottom) measurements from all observations between Galactic latitudes $10^{\circ} < |b| < 30^{\circ}$ (stars)  and $ |b| \geq 30^{\circ}$ (circles).
    In both spectra at lower $b$, an excess is present around the Fe complex---marked by the red arrows---a known feature of the GRXE but generally absent in CXB measurements, as evidenced by the $|b|\geq 30^\circ$ spectra.}
    \label{fig:specbcut}
\end{figure}

However, spectra generated from stacked images with this OBSID selection exhibit features beyond that expected for the CXB, namely the appearance of a line-like component around the same energies the Fe line complex was found to exist \citep[][]{Perez2019}.
The GRXE has a known emission line due to iron,
suggesting a stricter cut on galactic latitude was required \citep[see][Figure 4]{Revnivtsev2006_2}.
After some trials, the line feature was found to disappear when observations included only those with $|b| \geq 30^{\circ}$.
In Figure~\ref{fig:specbcut}, spectra are generated with this new criteria (circles) and compared to spectra made from observations with latitude $10^{\circ} < |b| < 30^{\circ}$ (stars).
In both FPMA (top panel) and FPMB (bottom panel), the line disappears with the stricter data selection (note that statistical quality of the black points is worse due to substantially lower total exposure times, 3~Ms when compared to the 8~Ms of the red points).
This stricter criteria reduces our overall exposure time by $\sim$17\% but limits continuum contributions from the GRXE, which are certainly also present in the spectrum but are less obvious as it is similar in shape to the CXB in this energy range.

Although observations with obvious stray light could be easily identified from the source detection step, artifacts in the initially stacked images revealed observations with more subtle stray or scattered light.
To ensure the observations included in stacks are free of faint, extended emission, images in the 3--20~keV band for each OBSID and FPM, were smoothed, and visually inspected.
These artifacts are more rare in observations at $|b| > 30^{\circ}$, where the bright source number per solid angle is lower than in the Galactic plane. This is consistent with the latest catalog of observations contaminated by stray light \citep{2021ApJ...909...30G}; we also cross-checked our clean OBSID list with that catalog.

\subsection{Solar Component}

\begin{figure}[h]
    \centering
    \includegraphics[scale=0.45]{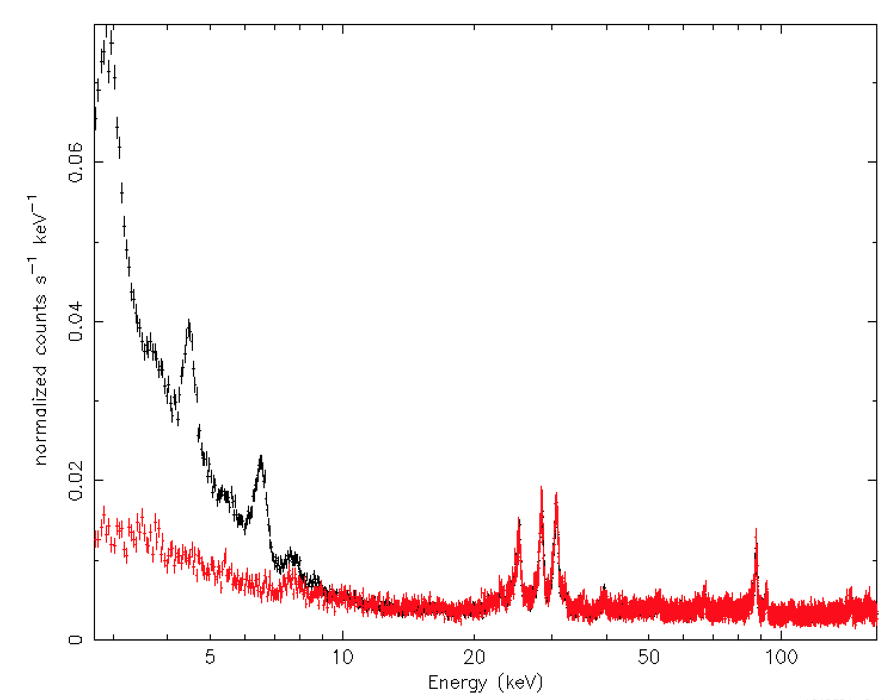}
    \caption{Spectra of stacked observations of Earth-occulted data from FPMB separated into periods when the observatory is illuminated by the Sun (black) or is in Earth's shadow (red).}
    \label{fig:02spectrum}
\end{figure}

\begin{figure}[h]
    \centering
    \includegraphics[width=\linewidth]{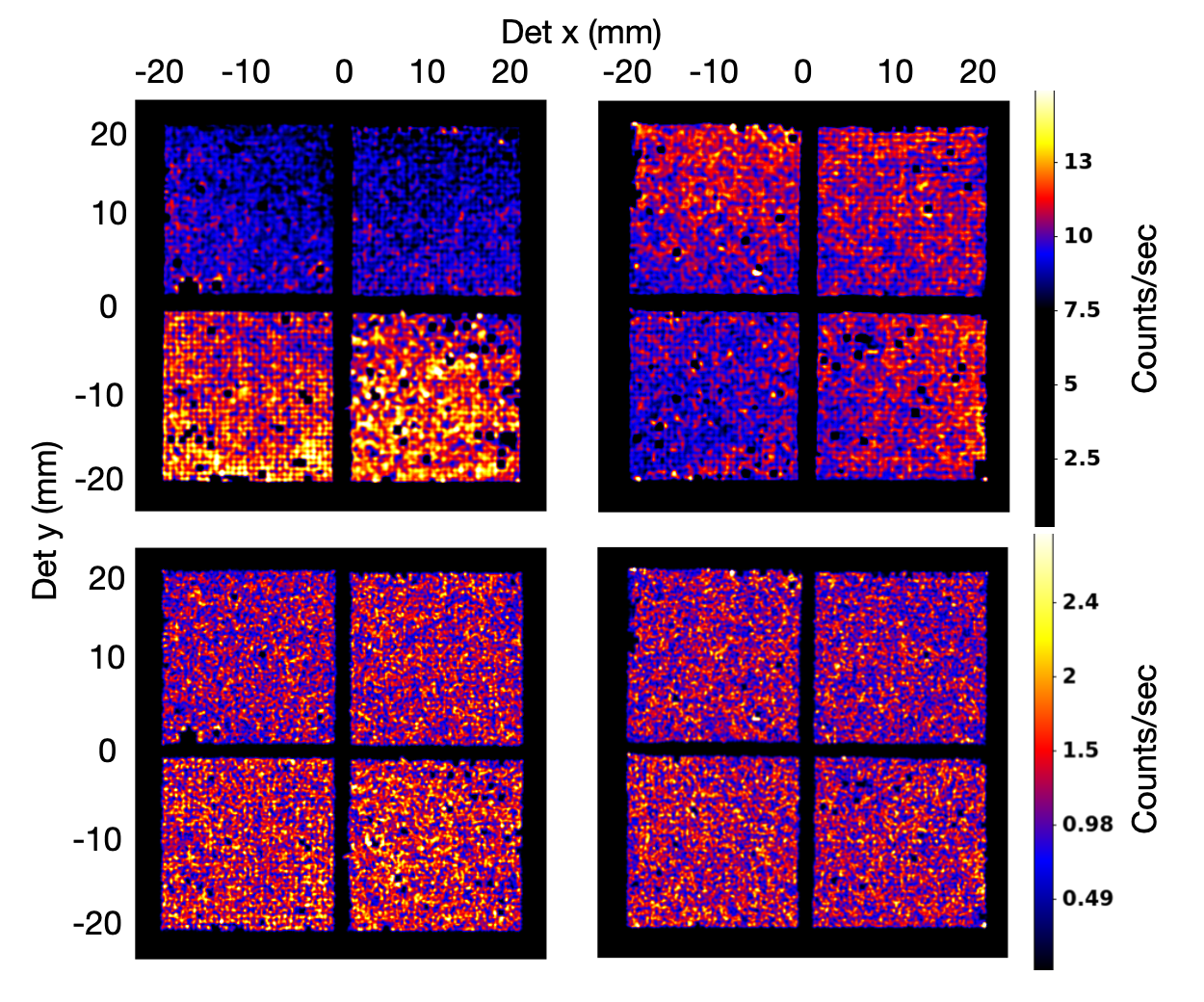}
    \caption{Stacked images from 3--8~keV of the Earth-occulted data for FPMA (left column) and FPMB (right column) during periods when \nustar\ is illuminated by the Sun (top row) and when it is in Earth's shadow (bottom row).
    A shadow pattern very similar to that produced by the CXB is apparent in the Sun-illuminated images but disappears when \nustar\ is in Earth's shadow.}
    \label{fig:02image}
\end{figure}

At energies below 8--10~keV, the background is generally dominated by solar activity. During a typical observation, the spacecraft will be exposed to direct and reflected radiation from the Sun, entering through the open optical path, 
possibly reflecting off the mast, optics bench, and backside of the aperture stops to reach the detectors. 
To show the impact of solar emission on \nustar's background, we compare stacked spectra from the entire FOV of FPMA and FPMB when the spacecraft points toward the Earth, filtered on whether \nustar\ was in sunlight or Earth's shadow (Fig.~\ref{fig:02spectrum} shows FPMB).
In addition, this enhanced solar component is not uniformly detected in the focal plane, but follows a pattern very similar to that produced by the aCXB (Fig.~\ref{fig:02image}).
We attempted to model this component, correlating its variation and estimating its flux from GOES monitoring observations, but these efforts were unsuccessful.
To avoid any confusion with the CXB signal we initially extract, we exclude all data when \nustar\ is illuminated by the Sun from our images, yielding a 50\% reduction in on-sky exposure time. 

During the initial image fitting after this cut, simply filtering on {\tt SUNSHINE} flag with {\tt nuscreen} did not completely eliminate the solar component,
though its presence in images was greatly reduced. 
Presuming the flag was not quite capturing the day/night transition, or that twilight periods were sufficient to cause the residual emission,
we culled an extra 600~s of exposure time per orbit where the flag changed, leading to an additional reduction in exposure time by 33\%.
This conservative cut eliminated any detectable solar signal and minimizes bias due to the Sun.
Even with these cuts, the statistical uncertainty on the measured 3--20~keV CXB flux is $\sim$2\%, comparable to or smaller than the systematic uncertainties.

\subsection{Detector Component}

\begin{figure}[ht]
    \centering
    \includegraphics[width=\linewidth]{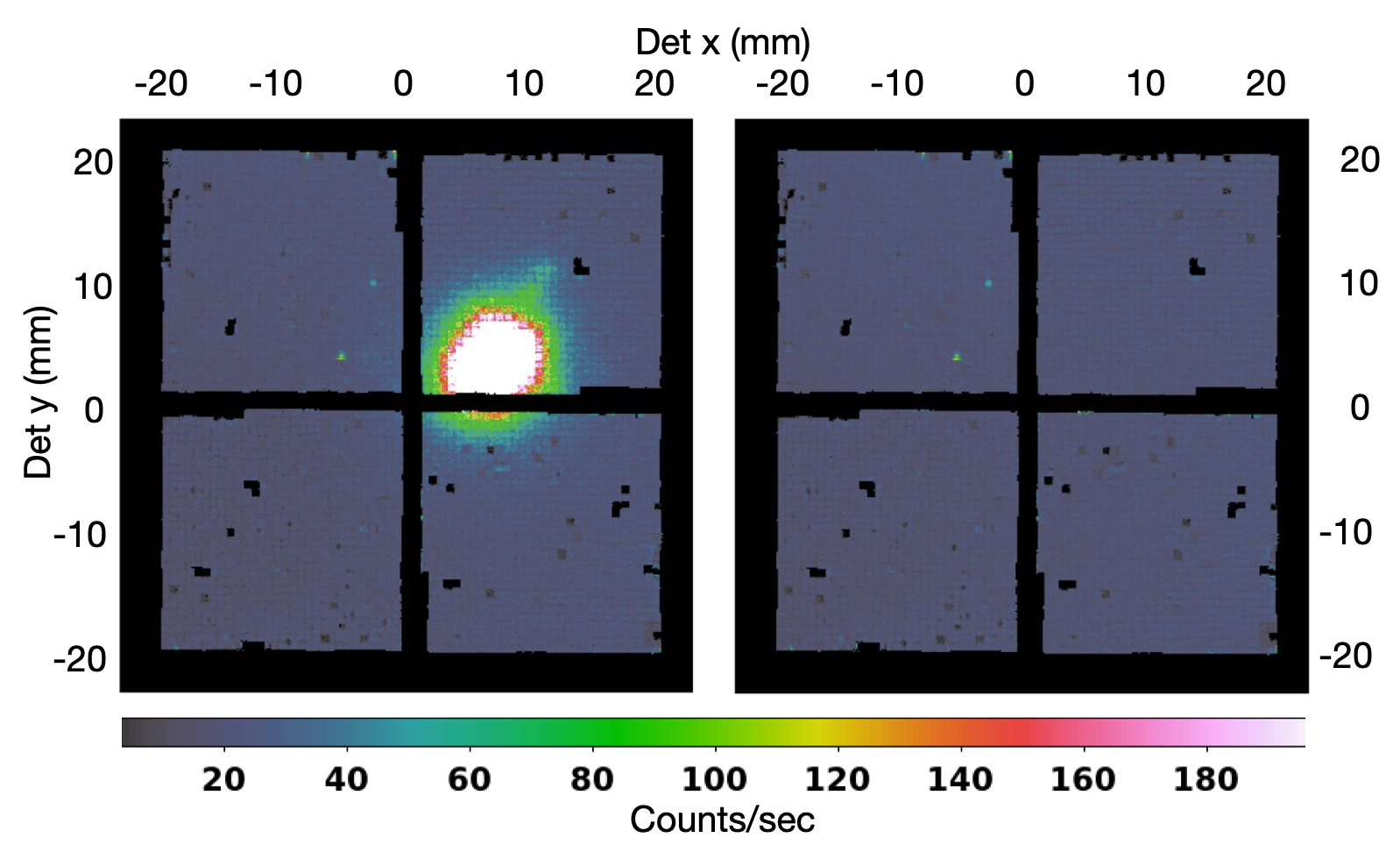}
    \caption{Earth-occulted stacked image for FPMB made from the default 02 {\tt nupipeline} filtering (left) and the same stacked image after applying an elevation limit of {\tt ELV}~$ < -3$\arcdeg\ with {\tt nuscreen} (right). The bright blob in the middle of the left image is the integrated sum of point source photons from target sources that were not completely obscured by the limb of the Earth due to the OCC data files starting to accept data before the limb is within the FOV.}
    \label{fig:lightleak}
\end{figure}

In a standard observation done by \nustar, there are two primary periods of data collection; one during which the telescope has an unobstructed view of the target, and one where the Earth is between the satellite and the target, completely obscuring the sky. Because no cosmic sources should be in the FOV due to the obstruction of the Earth, a clean measurement of the detector component, $I_c$, can be made.
However, the presence of a persistent bright spot near the optical axis location demonstrates that the default filtering criteria is insufficient---targets can still be seen, likely refracted through the atmosphere (Fig.~\ref{fig:lightleak}).
We therefore apply a more strict elevation criteria on our observations, {\tt ELEV}~$< -3$, to create Earth-occulted event files and images with {\tt nuscreen}, though the most recent calibration and analysis programs are able to remove all source photons at {\tt ELEV}~$< 0$.
With this criteria and by further selecting only times when the telescope is not illuminated by the Sun, we can isolate the instrumental background.
In Figure~\ref{fig:02image}, an aCXB-like spatial gradient due to the solar component dominates counts images at low energies (top panels), which disappear when Sun periods are excluded (bottom panels).
The instrumental component can be seen at all energies with a relatively flat spatial distribution, where the overall level between detectors vary by only small amounts.
We also performed a principal component analysis on the Earth-occulted, Sun-less images to discover any additional spatial features, but all major and minor principal components failed to indicate a significant shape besides flat.
We repeated this process for our full sky sampling, Figure~\ref{fig:mollweide} bottom, and our blank sky sampling, Figure~\ref{fig:mollweide} top.




\section{Results}
\label{sec:Results}

\begin{deluxetable}{lccc}
\tablecaption{Flux values measured for each set and subset of data for energies from 3 to 20 keV. The interval given for each value is the 90$\%$ confidence level. Units for flux are given as $10^{-11}$~erg~s$^{-1}$~cm$^{-2}$~deg$^{-2}$.  \label{tab:flux}}
\tablewidth{0pt}
\tablehead{
\colhead{DATA SET} & \colhead{FPMA} & \colhead{FPMB} & \colhead{TIED\tablenotemark{a}} \\
\colhead{} & \colhead{Flux$\rm{_{3-20keV}^{CXB}}$} & \colhead{Flux$\rm{_{3-20keV}^{CXB}}$} & \colhead{Flux$\rm{_{3-20keV}^{CXB}}$} 
}
\startdata
  FULL SET$\tablenotemark{b}$  & 2.41$_{-0.03}^{+0.04}$ & 2.56$_{-0.01}^{+0.01}$ & 2.46$_{-0.10}^{+0.10}$ \\
  Blank Sky$\tablenotemark{c}$ & 2.38$_{-0.05}^{+0.06}$ & 2.53$_{-0.03}^{+0.03}$ & 2.43$_{-0.03}^{+0.02}$ \\
  $b>0^{\circ}$;$l<180^{\circ}$ $\tablenotemark{*}$       & 2.31$_{-0.04}^{+0.04}$ & 2.51$_{-0.06}^{+0.06}$ & 2.44$_{-0.05}^{+0.06}$ \\
  $b>0^{\circ}$;$l>180^{\circ}$ $\tablenotemark{*}$    & 2.28$_{-0.08}^{+0.07}$ & 2.48$_{-0.03}^{+0.04}$ & 2.39$_{-0.05}^{+0.06}$ \\
  $b<0^{\circ}$;$l<180^{\circ}$ $\tablenotemark{*}$       & 2.10$_{-0.04}^{+0.04}$ & 2.60$_{-0.09}^{+0.10}$ & 2.40$_{-0.06}^{+0.05}$ \\
  $b<0^{\circ}$;$l>180^{\circ}$ $\tablenotemark{*}$     & 2.21$_{-0.08}^{+0.08}$ & 2.62$_{-0.08}^{+0.09}$ & 2.45$_{-0.11}^{+0.05}$
\enddata
\footnotesize{$^{a}$Photon index, cutoff energy, and norm values  for FPMA and FPMB are tied together for a joint fit. $^{b}$Utilizes all observations as represented in Figure~\ref{fig:mollweide}. $^{c}$Subset of the full set, this data set are all observations of the full that have no observable sources as represented in Figure~\ref{fig:mollweide}. $^{*}$~Quadrant limits of the sky in galactic coordinates that this subset of data was analyzed in. i.e., all observations from $b>0^{\circ}$ and $l>180^{\circ}$ is the part of sky where the COSMOS field is found.}

\end{deluxetable}

\subsection{Final Values}
\begin{figure}[t]
    \centering
    \includegraphics[width=\linewidth]{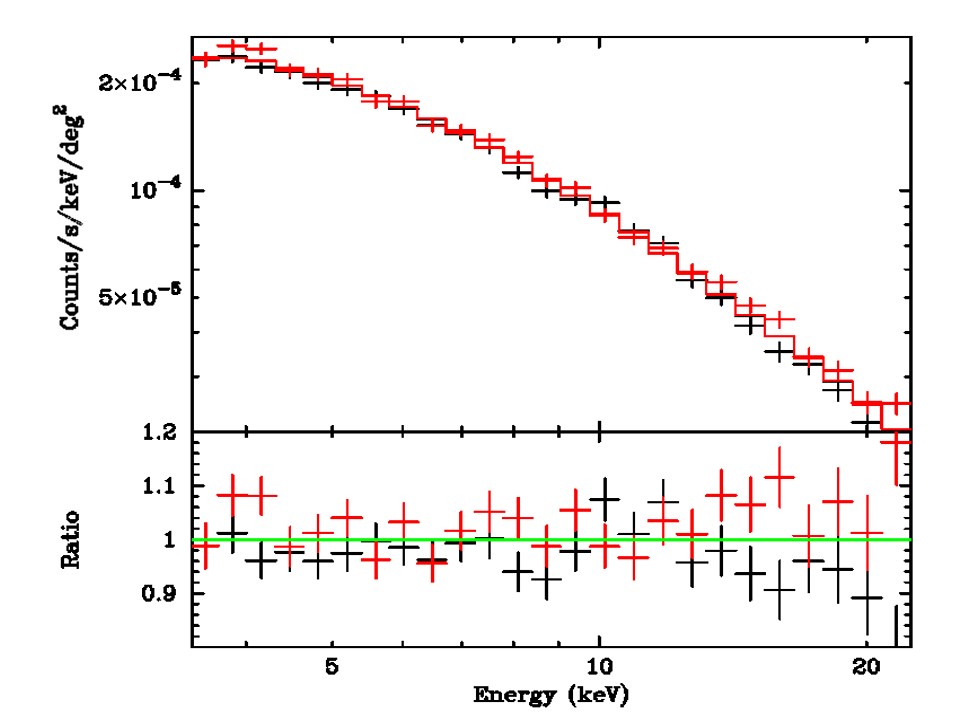}
    \includegraphics[width=\linewidth]{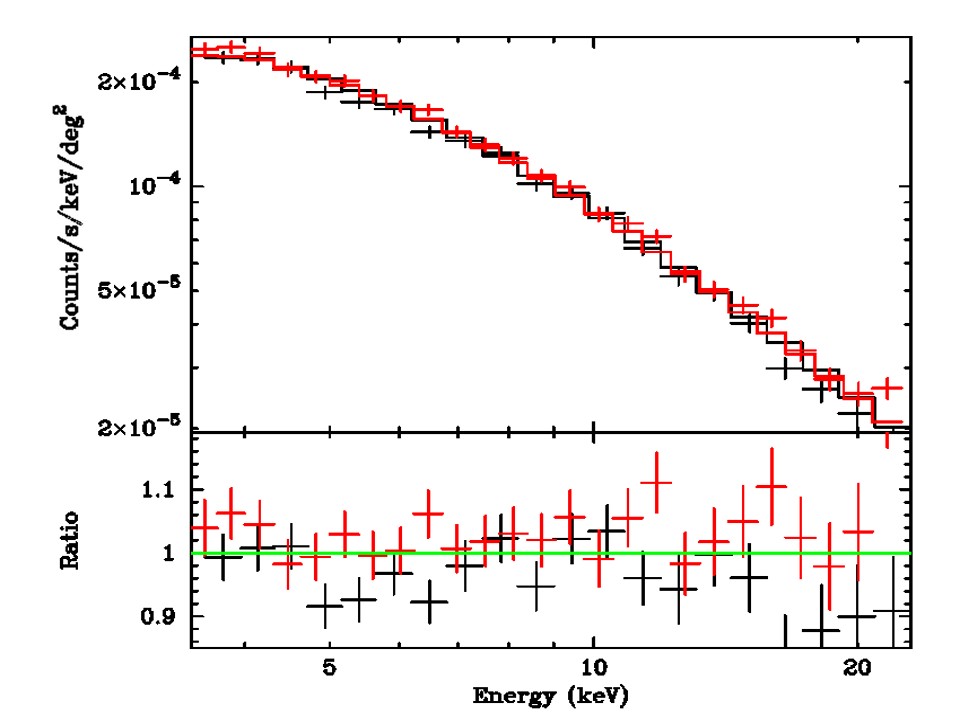}
    \caption{Measurements of the aCXB component extracted from spatial fits in narrow energy bands for FPMA (black) and FPMB (red) of the Blank Sky list (top panel) and the Full Sky list (bottom panel), jointly fit with the cutoff power law model.
    For each observation set, the upper panel shows the measurements and best-fit model, in units of counts~s$^{-1}$~keV$^{-1}$~deg$^{-2}$, and the lower panel gives the ratio of the data to the model.
    }
    \label{fig:cxbspec}
\end{figure}

Spectral fitting was applied to both the blank sky (top panel) and full sky (bottom panel) measurements, shown in Figure~\ref{fig:cxbspec}, and to four quadrants of the sky divided by $|b| = 0^{\circ}$ and $|l| = 180^{\circ}$ in Galactic coordinates; flux values for these fittings are in Table~\ref{tab:flux}. 
The cutoff power law ({\tt cutoffpl} in {\tt XSpec}) model was used in each of these spectra and FPMA and FPMB were independently and jointly fit in the 3--20~keV energy range. 
We found photon indices of $\Gamma \approx 1.4$ for both FPMA and FPMB, while the cutoff energy differed between the two instruments as can be seen in Table~\ref{table:fitparams}. In order to directly compare our measurements with those from \heao\ \citep[][]{Revnivtsev2005--HEAO1-cxb,Jahoda-2005-HEAO-CXB} at softer energies, we limited our energy range to 3--10~keV and fit with a simple power law model (see Table~\ref{table:fitparams}). The reported photon indices from \heao\ were 1.4 and 1.558, respectively, and we measured a value of $\Gamma \approx 1.55\pm0.04$ when the FPMs were fit jointly for the NOSUN data sets. For the full exposure data sets, we find $\Gamma \approx 1.52\pm0.02$. Finally, we applied the model described in \citet[][]{Gruber99} as a {\tt cutoffpl} with the photon index fixed at $\Gamma_{G99} = 1.29$ with the cutoff energy $E_c$ first allowed to be free, and then fixing its value to $E_c = 41.13$~keV and allowing the photon index to be free. Our best-fit values, $E_c = 23.20\pm1.54$~keV and $\Gamma = 1.44\pm0.02$, in each case respectively, are not consistent with the \citet{Gruber99} model, driven by the lower flux measured by \nustar\ at higher energies. When we fix $E_c$ to the value found for the energy range of 3--20~keV, we found a joint FPMA/FPMB value of $\Gamma \approx 1.43\pm0.09$ for the NOSUN data set and $\Gamma \approx 1.38\pm0.02$ for the full exposure set. Both sets found similar cutoff energies of $E_c \approx 37.3\pm1.6$~keV for the NOSUN data set and $E_c \approx 37.9\pm2.5$ for the full exposure set.

We also fit spectra only in the high energy 10--20~keV bandpass with the
{\tt cutoffpl} and {\tt powerlaw} models for comparison, see Table~\ref{table:fitparams}. Photon indices were found to be higher for the {\tt powerlaw} fits than those of the {\tt cutoffpl}, which have values more consistent with those found in the lower energy 3--10~keV bandpass. 
The energy binning above 10~keV becomes necessarily wider due to a decrease in count rate; this means the number of energy bins in this energy range are less than those found below 10~keV, giving less data points for model fits. The increase in exposure time for the full data set had significant influence on FPMA as seen by the increase in the cutoff energy, more consistent photon indices, and similar fluxes as those found with FPMB.


Flux measurements for the NOSUN data set in energies from 3--20~keV were done for the entire survey of observations (Full Sky list), all observations where an exclusion region due to sources was not found (Blank Sky list), and for all observations in each quadrant of the sky in Galactic coordinates that coincides with a large survey field which is used as a reference for that quadrant (see Table~\ref{tab:flux}). Flux measurements were done in the 3--10~keV, 10--20~keV, and 3--20~keV energy ranges for each model, applied as mentioned above for both the NOSUN data set and the full exposure data set. Fluxes and their errors were calculated in {\tt XSpec} with the {\tt cflux} convolution model; uncertainties are reported at the 90\% confidence level.

Due to the $\sim10\%$ difference between our measurement and that of the pilot study \citep{2021MNRAS.502.3966K}, we measured the flux in the survey fields used in that study filtered in a similar way: without excluding periods where the spacecraft is illuminated by the Sun. As was done with the previous study, a low energy flux correction in the form of a broken power law ({\tt bknpower}) was applied in conjunction with the {\tt cutoffpl} model. We used the parameters given in the study to model the solar emission ({\tt bknpower}, $\Gamma_1 = 5.0$, $E_c = 4.8$, and $\Gamma_2 = 0.9$), and attempted to fit the {\tt cutoffpl} model that describes the CXB ($\Gamma = 1.29$ and $E_c = 41.13$) with only the normalizations allowed to be free. Our reduced chi-squared fit statistic was $\chi_{\rm red}^2 > 2.0$ and our flux value was found to be $\sim 10\%$~lower than what was found in the previous study. The parameters of the {\tt cutoffpl} were then set to the values found in this study, $\Gamma \approx 1.39\pm0.3$ and $E_c \approx 40.0\pm2.0$~keV, with the normalization allowed to be free. The reduced chi-squared value was found to be $\chi_{\rm red}^2 = 1.2$ with an averaged flux of $2.53\pm0.04\times 10^{-11}$~erg~s$^{-1}$~cm$^{-2}$~deg$^{-2}$ in the 3--20~keV bandpass. Our fluxes, when limited to energies $\geq 10$~keV, were more in line with the $10\%$~higher flux of the previous study with adjusted values of $2.77\pm0.08\times 10^{-11}$~erg~s$^{-1}$~cm$^{-2}$~deg$^{-2}$ in the energy range of 3--20~keV.

\begin{figure}[t]
    \centering
    \includegraphics[width=\linewidth]{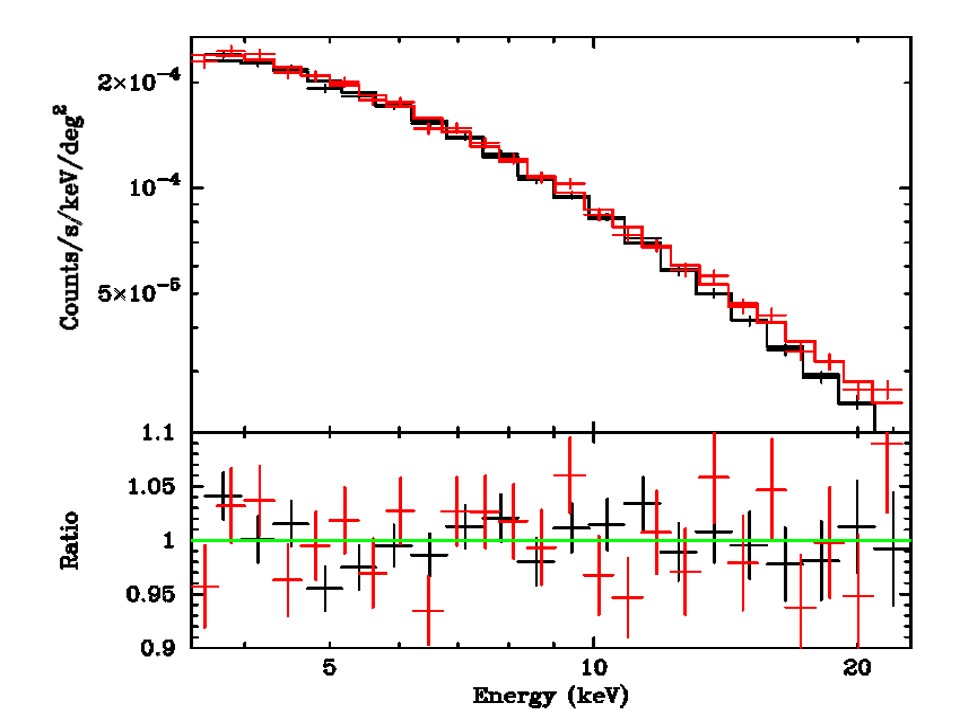}
    \caption{Same as Fig.~\ref{fig:cxbspec} but for the full observational catalog with the full exposure time, $\sim18$~Ms per telescope. The increase in exposure time provides better agreement between FPMA and FPMB. The fluxes in Table~\ref{table:fitparams} for FPMB are similar to those reported for FPMB in the NOSUN data sets.
    }
    \label{fig:A_B_fullspectra}
\end{figure}

In an effort to reach better agreement between the reported results from FPMA and FPMB for the aCXB, we tied the detector components $a_0$ through $a_3$ to each other using ratios determined from Earth-occulted observations. This application is possible due to the well characterized and expected responses of each detector plane \citep[][]{Kitaguchi-2011SPIE.8145E..07K}. These ratios were found from fits to the OCC NOSUN data, which provides the most accurate representation of the detector-only signal in any observation. These ratios were then measured and adjusted for exposure time and energy bin width. Setting $a_0$ as the reference, the other 3 normalizations were determined within the aCXB optimization function as a function of the relative variance that detector has compared to $a_0$. Further, a hard upper limit for $a_0$ was set by the exposure adjusted value found within the OCC NOSUN data set. These constraints had little impact on the overall fluxes below 10~keV for either FPMA or FPMB; however, above 10~keV FPMB become more consistent with the \heao\ measurement, while FPMA saw little overall change in its flux measurements. From this analysis a difference between FPMA and FPMB, on average for energies $>10$~keV, decreased by about $5\%$. Finally, $a_3$ on FPMA was excluded from the analysis due to its consistently higher normalization value given in all analysis, thought to be associated with the detector thickness variation. FPMA $a_3$ has also been seen to exhibit larger variations in signal then other detectors, as can be seen in figure~\ref{fig:02image}. 
It was found, and can be seen in figures~\ref{fig:cxbspec} and figure~\ref{fig:A_B_fullspectra}, that the energy binning for FPMA had to be larger than what could be achieved with FPMB and still report consistent values for the aCXB. This was verified by increasing the energy bin widths for both telescopes, which returned similar values as those reported in this study. While both telescopes have similar count rates for their spectra, this approach leaves FPMA with fewer data points and smaller errors while FPMB has more data points and slightly larger errors.

Our flux measurement, when compared to previous measurements in this energy range, is more in agreement with \heao\ up to 10~keV. \heao\ was measured at $F_{\rm{3-20keV}}^{G99} =$ 2.61 $\times 10^{-11}$~erg~s$^{-1}$~cm$^{-2}$~deg$^{-2}$. When we limited our energy to 3--10~keV in our NOSUN data set, our flux had a measurement of $2.46 \times 10^{-11}$~erg~s$^{-1}$~cm$^{-2}$~deg$^{-2}$ for telescope A and $2.59 \times 10^{-11}$~erg~s$^{-1}$~cm$^{-2}$~deg$^{-2}$ for telescope B with a combined flux measurement of $2.53 \times 10^{-11}$~erg~s$^{-1}$~cm$^{-2}$~deg$^{-2}$ -- with values reported in the energy range of 3--20~keV. When this same analysis is applied to the full exposure set, the measurements for both FPMA and FPMB fall within $\sim2\%$~of flux measurements in the 3--10~keV range. When adjusted to 3--20~keV, the flux is found to be $\sim 2.59 \times 10^{-11}$~erg~s$^{-1}$~cm$^{-2}$~deg$^{-2}$. After tying the detectors and excluding detector 3 on FPMA our flux measurements for the energy range of 3--20~keV become $2.63 \times 10^{-11}$~erg~s$^{-1}$~cm$^{-2}$~deg$^{-2}$ for telescope A and $2.58 \times 10^{-11}$~erg~s$^{-1}$~cm$^{-2}$~deg$^{-2}$ for telescope B with a combined flux measurement of $2.61 \times 10^{-11}$~erg~s$^{-1}$~cm$^{-2}$~deg$^{-2}$.

\subsection{Systematic Errors}


The stray light pattern maps that model the expected CXB on the FPMs are created by careful determination of the interplay between the FOV of the sky around the optics bench each FPM sees and the section of the sky each aperture stop blocks. Verification in the absolute measurement of the optics bench shape, while believed to be highly accurate, could not be done, creating a possibility for a systematic offset in relation to the FPM gradient maps due to this needed precision in their creation. Further, there is a known degeneracy in relation to the gradient of these maps that favors the lowest value along that gradient when fitting with $\chi^2$ statistic. As stated, when we allowed the gradient to shift in a limited parameter space around the optimal position, we found the systematic error was small --- $\sim1\%$.

From Table~\ref{tab:flux}, there is a systematic $\sim 5\%$ difference in flux measured between FPMA and FPMB for the NOSUN data set. While it is common to see statistical variation between two measurements around a common value, we show in every measurement a higher flux value in B when compared to A, suggesting a systematic issue. Limits in energy band fitting, energy band width, and positional shifting of the aCXB gradient model to allow flexibility in fitting, as stated above, changed the flux by $\sim 1\%$. This nominal variation was also observed when centering the aCXB gradient at the most unfavorable gradient position when compared to the optimal position in our 16$\times$16 exploratory grid space. The increase in exposure time in the full exposure data set lessens the difference between FPMA and FPMB significantly and further suggests that this systematic error is primarily motivated by overall count statistics and noise variations associated with FPMA. When detector values were tied and an improvement to the FPMB measurement for energies above 10~keV was identified, FPMA continued to show previous type values and only further disagreed with FPMB. Only after excluding {\tt DET3} on FPMA was the measurement between the two telescopes consistent with each other, further suggesting that the variations seen in {\tt DET3} and partially in {\tt DET2} of FPMA, bias the detector normalization in those detectors higher than expectation, and thus suppressing the overall aCXB normalization value for this telescope. While on the surface this issue has been resolved, it is, as of yet, unknown which detector on FPMA has the most significant contribution to the detector normalization and still can be considered a source of systematic error.

The Crab nebula flux measurement, Figure~\ref{fig:crab}, had a $1.5\%$ variance when measured from 3--22~keV from the referenced work showing another source of error that increases our uncertainty within our measurement. When we extended our energy range to 40~keV, this variance increased to $3.6\%$. These results are not thought to be associated with the systematic issue between telescope A and B mentioned above and is considered a separate error due to the change in the shape of the stray light gradient due to the transparency of the aperture stop in energies above 20~keV, the selection of analysis area that may differ from the original work, or the selection of pixels, specifically those that are excluded in this work such as all edge pixels and those that were deemed problematic as their high, consistent count rate would bias any measurement.

\begin{deluxetable*}{cccccc}
\tablecaption{Fit parameters used for models and energies given.  \label{tab:params}}
\tablewidth{0pt}
\tablehead{
\colhead{Energy} & \colhead{Model$^\alpha$} & \colhead{$\Gamma$} & \colhead{$E_{\rm c}$} & \colhead{Flux$^{\dagger}$} & \colhead{$\chi^2$/dof}
}
\startdata
\multicolumn{6}{c}{FPMA} \\
    3--20~keV & CUTOFFPL$^{\alpha}$ & $1.39\pm0.14$ & $29.6\pm1.6$ & $2.41\pm0.04$ & 26.29/17\\
    3--10~keV & PL$^{\beta}$ & $1.53\pm0.06$ & -- & $1.33\pm0.01$ & 10.75/11 \\
    3--10~keV & CUTOFFPL$^{\alpha}$ & $1.38\pm0.05$ & 29.6$^{*}$ & $1.35\pm0.05$ & 16.64/11 \\
    10--20~keV & PL$^{\beta}$ & $2.09\pm0.16$ & -- & $1.02\pm0.07$ & 3.94/4 \\
    10--20~keV & CUTOFFPL$^{\alpha}$ & $1.55\pm0.21$ & $28.5\pm2.1$ & $1.02\pm0.07$ & 3.96/3\\
    \multicolumn{6}{c}{FPMB} \\
    3--20~keV & CUTOFFPL$^{\alpha}$ & $1.44\pm0.12$ & $44.0\pm2.8$ & $2.56\pm0.01$ & 21.25/22 \\
    3--10~keV & PL$^{\beta}$ & $1.56\pm0.05$ &  -- & $1.39\pm0.02$ & 12.22/14 \\
    3--10~keV & CUTOFFPL$^{\alpha}$ & $1.45\pm0.05$ & 44.0$^{*}$ & $1.41\pm0.05$ & 13.86/14 \\
    10--20~keV & PL$^{\beta}$ & $1.94\pm0.16$ & -- & $1.12\pm0.06$ & 5.59/6 \\
    10--20~keV & CUTOFFPL$^{\alpha}$ & $1.44\pm0.15$ & $44.5\pm4.3$ & $1.10\pm0.08$ & 5.90/5 \\
    \multicolumn{6}{c}{TIED} \\
    3--20~keV & CUTOFFPL$^{\alpha}$ & $1.43\pm0.09$ & $37.3\pm1.6$ & $2.46\pm0.10$ & 70.54/42 \\
    3--10~keV & PL$^{\beta}$ & $1.55\pm0.04$ & -- & $1.36\pm0.01$ & 40.11/27\\
    3--10~keV & CUTOFFPL$^{\alpha}$ &  $1.42\pm0.03$ & 37.3$^{*}$ & $1.38\pm0.06$ & 46.91/27\\
    10--20~keV & PL$^{\beta}$ & $1.99\pm0.11$ & -- & $1.07\pm0.04$ & 22.28/12\\
    10--20~keV & CUTOFFPL$^{\alpha}$ & $1.41\pm0.15$ & $24.3\pm0.99$ & $1.07\pm0.10$ & 21.71/11\\ \hline
    \hline
    \multicolumn{6}{c}{Full Exposure Time} \\ \hline
    \multicolumn{6}{c}{FPMA} \\
    3--20~keV & CUTOFFPL$^{\alpha}$ & $1.39\pm0.02$ & $38.74\pm2.74$ & $2.51\pm0.03$ & 21.70/17\\
    3--10~keV & PL$^{\beta}$ & $1.52\pm0.03$ & -- & $1.37\pm0.01$ & 9.97/11 \\
    3--10~keV & CUTOFFPL$^{\alpha}$ & $1.36\pm0.03$ & 38.7$^{*}$ & $1.38\pm0.01$ & 11.37/11 \\
    10--20~keV & PL$^{\beta}$ & $1.99\pm0.12$ & -- & $1.10\pm0.01$ & 2.87/4 \\
    10--20~keV & CUTOFFPL$^{\alpha}$ & $1.62\pm0.02$ & $38.7\pm3.11$ & $1.10\pm0.01$ & 2.10/3\\
    \multicolumn{6}{c}{FPMB} \\
    3--20~keV & CUTOFFPL$^{\alpha}$ & $1.38\pm0.03$ & $39.1\pm3.0$ & $2.58\pm0.03$ & 29.7/22 \\
    3--10~keV & PL$^{\beta}$ & $1.52\pm0.03$ &  -- & $1.40\pm0.01$ & 15.89/14 \\
    3--10~keV & CUTOFFPL$^{\alpha}$ & $1.37\pm0.04$ & 39.1$^{*}$ & $1.40\pm0.01$ & 17.51/14 \\
    10--20~keV & PL$^{\beta}$ & $1.79\pm0.18$ & -- & $1.13\pm0.03$ & 7.76/6 \\
    10--20~keV & CUTOFFPL$^{\alpha}$ & $1.40\pm0.18$ & $37.0\pm1.7$ & $1.13\pm0.02$ & 6.45/5 \\
    \multicolumn{6}{c}{TIED} \\
    3--20~keV & CUTOFFPL$^{\alpha}$ & $1.38\pm0.02$ & $37.9\pm2.5$ & $2.54\pm0.03$ & 67.16/42 \\
    3--10~keV & PL$^{\beta}$ & $1.52\pm0.02$ & -- & $1.38\pm0.01$ & 37.07/27\\
    3--10~keV & CUTOFFPL$^{\alpha}$ &  $1.36\pm0.02$ & 37.9$^{*}$ & $1.39\pm0.01$ & 41.77/27\\
    10--20~keV & PL$^{\beta}$ & $1.88\pm0.15$ & -- & $1.12\pm0.02$ & 22.33/12\\
    10--20~keV & CUTOFFPL$^{\alpha}$ & $1.50\pm0.18$ & $37.0\pm0.10$ & $1.12\pm0.10$ & 20.46/11\\ \hline
\enddata
\footnotesize{ ~ $^{\dagger}$ Calculated for the energy range of each fit (3--20, 3--10, and 10--20 keV) and given in units of $10^{-11}$~erg~s$^{-1}$~cm$^{-2}$~deg$^{-2}$ \\
$^{\alpha}$ CUTPL model: $I(E) \propto E^{-\Gamma}\exp(-E/E_{\rm c})$; 
PL model: $I(E) \propto E^{-\Gamma}$ \\ 
$^{*}$ $E_{\rm c}$ in the 3--10~keV fits were fixed to the best-fit from the 3--20~keV fits
}
\label{table:fitparams}
\end{deluxetable*}

 
\begin{deluxetable*}{cccccc}
\tablecaption{Fit parameters for FPMA and FPMB when the detector norm values were tied and limited by a ceiling value. In addition, FPMA is measured while excluding detector 3 (DET3).  \label{tab:params}}
\tablewidth{0pt}
\tablehead{
\colhead{Energy} & \colhead{Model$^\alpha$} & \colhead{$\Gamma$} & \colhead{$E_{\rm c}$} & \colhead{Flux$^{\dagger}$} & \colhead{$\chi^2$/dof}
}
\startdata
\multicolumn{6}{c}{FPMA} \\
    3--20~keV & CUTOFFPL$^{\alpha}$ & $1.36\pm0.05$ & $38.4\pm7.2$ & $2.63\pm0.05$ & 15.95/15 \\
    10--20~keV & PL$^{\beta}$ & $1.86\pm0.13$ & -- & $1.19\pm0.08$ & 2.25/3 \\
    10--20~keV & CUTOFFPL$^{\alpha}$ & $1.61\pm0.28$ & $44.7\pm9.9$ & $1.19\pm0.13$ & 2.19/2 \\
\multicolumn{6}{c}{FPMB} \\
    3--20~keV & CUTOFFPL$^{\alpha}$ & $1.38\pm0.02$ & $40.58\pm2.6$ & $2.58\pm0.02$ & 27.31/22\\
    10--20~keV & PL$^{\beta}$ & $1.67\pm0.01$ & -- & $1.15\pm0.02$ & 7.71/6\\
    10--20~keV & CUTOFFPL$^{\alpha}$ & $1.35\pm0.01$ & $40.1\pm2.6$ & $1.15\pm0.01$ & 6.66/5\\ 
\multicolumn{6}{c}{TIED}  \\
    3--20~keV & CUTOFFPL$^{\alpha}$ & $1.38\pm0.04$ & $38.1\pm3.9$ & $2.61\pm0.03$ & 41.80/40\\
    10--20~keV & PL$^{\beta}$ & $1.89\pm0.07$ & -- & $1.16\pm0.04$ & 8.04/10\\
    10--20~keV & CUTOFFPL$^{\alpha}$ & $1.61\pm0.13$ & $40.5\pm7.3$ & $1.16\pm0.09$ & 10.2/9\\ \hline
\enddata
\footnotesize{ ~ $^{\dagger}$ Calculated for the energy range of each fit (3--20 and 10--20 keV) and given in units of $10^{-11}$~erg~s$^{-1}$~cm$^{-2}$~deg$^{-2}$ \\
$^{\alpha}$ CUTPL model: $I(E) \propto E^{-\Gamma}\exp(-E/E_{\rm c})$; 
PL model: $I(E) \propto E^{-\Gamma}$ \\ 
}
\tablecomments{This data set reflects the model parameters for the CXB normalization measurement when the detector normalization values were tied together with a maximum value estimated from the occulted data set when the satellite was in Earth's shadow. FPMA showed little to no change when limited to only tying the detector values to $a_0$ over the range of energies with fit parameter values that were within $1\%$ of the given values in Table~\ref{table:fitparams}. After the exclusion of detector $a_3$ on FPMA, measurement values were found to be more consistent between both telescopes. For energies of 3--10~keV, the fit parameters for FPMA and FPMB followed what was reported in Table~\ref{table:fitparams} so no new measurement is reported here.}

\label{table:FPMBonly}
\end{deluxetable*}


\section{Summary and Discussion}
\label{sec:Summary}

\begin{figure*}
    \centering
    \includegraphics[width=\textwidth]{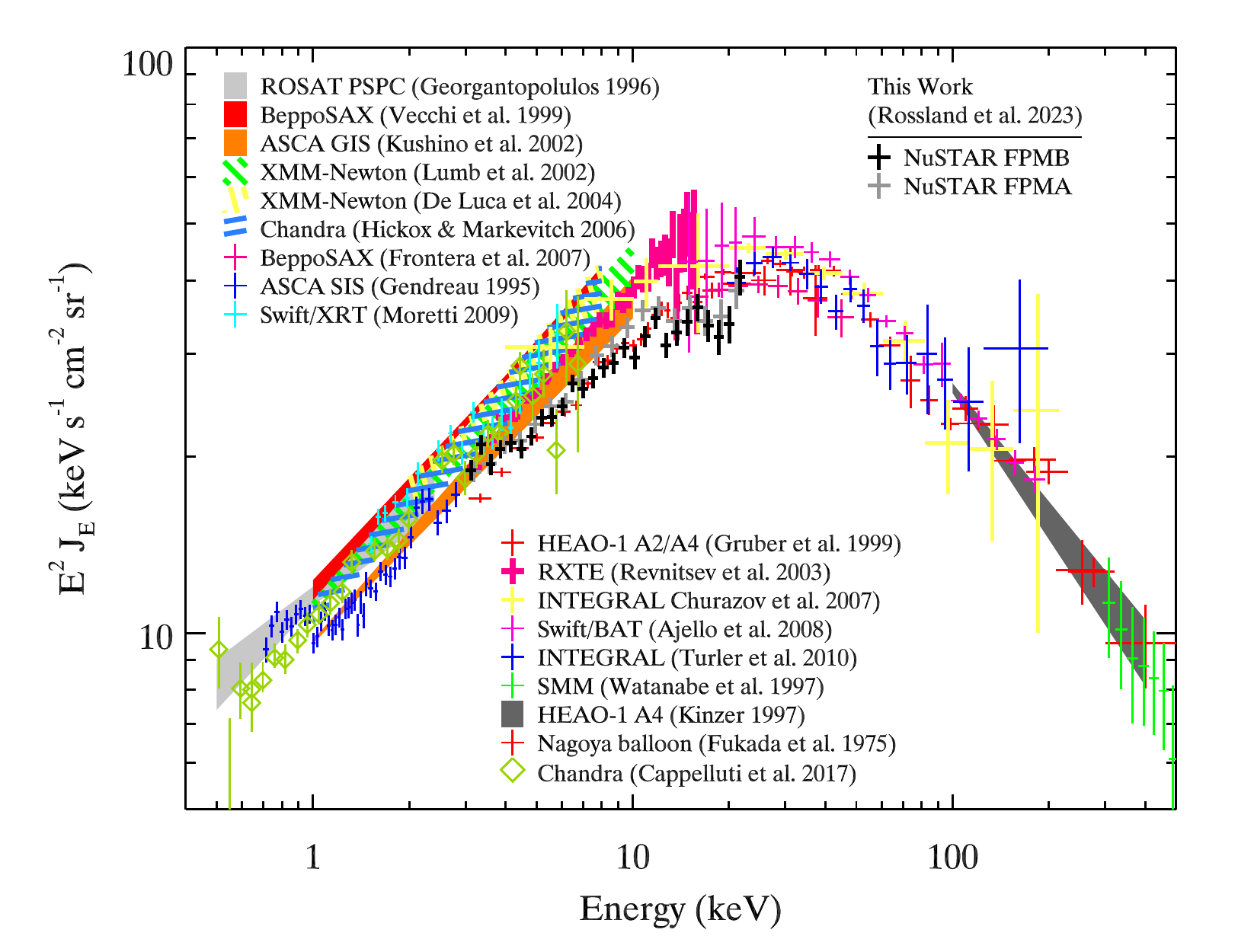}
    \caption{The energy spectrum of the cosmic X-ray background as measured in past studies and by \nustar\ in this work, using the best-fit cut-off power law model found from joint fits to FPMA (grey crosses) and FPMB (black crosses). Detector components $a_0$ through $a_3$ were tied together and limited to a maximum value based on the Earth-occulted NOSUN measurements and detector $a_3$ on FPMA was excluded, adjusted for energy band width and exposure time.
    Below 10~keV, the \nustar\ measurements agree very well with those from the \heao\ A2/A4 analysis (partly obscuring their thin red crosses). Above 10~keV FPMA and FPMB still largely agree with \heao\, however the aCXB starts to become less dominant then other background compoenents in \nustar\ and become strongly influanced by strong emission lines found at these higher energies.
    \citep[Past CXB measurements are taken from][as indicated in the plot.]{Fukada-1975Natur.254..398F,Gendreau1995-ASCA,Georgantopoulos-1993MNRAS.262..619G,Kinzer1997-heao-1A4,Watanabe-1997AIPC..410.1223W,Gruber99,Vecchi-1999A&A...349L..73V,Lumb2002-XMM,Kushino-2002PASJ...54..327K,Revnivtsev-2003A&A...411..329R,DeLuca2004-XMM,2006ApJ...645...95H,Frontera2007-BeppoSAX,Churazov2007-integral,Ajello2008-swift,Moretti2009-swift/xrt,Tuler2010-integral,Cappelluti-2017}
    }
    \label{fig:cxb324}
\end{figure*}

The first mission-specific measurement of the CXB was made over 40 years ago on a continuously surveying space-based platform, \heao, including the A2 instrument, with six 
proportional counters covering energies from 0.2--60~keV. Subsequent measurements of the CXB by more recent missions generally report fluxes $\geq10\%$~higher than the \heao\ A2 measurement, typically with a flatter slope.
Figure~\ref{fig:cxb324} presents the \nustar\ measurements of the CXB along with previous measurements by other X-ray observatories.
Absolute calibration at X-ray energies is a notorious problem given the lack of non-variable, bright point sources, resulting in the historical use of mildly extended and variable sources like the Crab nebula/pulsar.
Below 10~keV, most missions employed focusing optics with detectors that often require special operating modes to observe calibration sources.
At higher energies, coded mask techniques are used, which are challenged by the higher source surface density at the lower energy end of their bandpasses.
\nustar\ bridges these types of missions in both energy and technique.
In this study, we use \nustar's aperture light leak as a simplistic type of coded mask with detectors that operate identically for bright and faint sources, allowing a measurement of the CXB most like that of \heao\ but with more modern technology.

Of the $\sim$1400 scientific observations per telescope originally selected from 2012--2017 archival data, $\sim$700 
met our initial requirements. From those, only $\sim$600 observations per telescope were used in our final survey. The initial preparation of our survey data was performed to ensure maximum retention of exposure time per observation by developing an automatic procedure to identify and remove flaring events. Later, in an effort to remove all emission caused by the Sun, we trimmed exposure time from the ends of observing windows to avoid periods of twilight. This allowed us to make a CXB measurement with minimal influence from known nuisance sources and gave us the framework to estimate the reliability of CXB measurements made when those sources might interfere.

The systematic difference between the NOSUN measurement of FPMA and FPMB in our final measurements was unexpected. Our flux measurements of the Crab and of the CXB using blank survey fields with similar data selection as  \citet{2021MNRAS.502.3966K} gave self-consistent estimates for FPMA and FPMB within the given statistical uncertainties with FPMB consistently returning a higher flux value. 
Removing periods of Sun illumination, given its generally steeper spectral shape than the CXB, should not have preferentially affected one instrument over the other.
When we analyzed the full time data set, and allowed FPMA to have larger energy bins, we found it drastically lessened the difference between the two telescopes, suggesting we might be at a measurable limit for FPMA, since these variations only affected that telescope.
Another possibility is that there is a slight degeneracy between the aCXB gradient and the relative normalization of the instrumental background in one of the instruments.
For FPMA, we found higher instrumental normalizations for {\tt DET2} and {\tt DET3} compared to those of {\tt DET0} and {\tt DET1}. While this is expected due to the shape of the aCXB gradient on FPMA, these values were beyond that expectation.
This trend broadly matches the aCXB shape (Fig.~\ref{fig:grads}) and allows for the possibility of confusion, with a fraction of CXB emission being mistakenly modeled by the instrumental component.
That this confusion isn't seen in other measurements may be attributable to the spatial gradient caused by the Sun's emission on the detector at energies below $\sim 10$~keV. Its shape, when integrated over large exposure times, 
is similar, if not exactly identical, to the aCXB shape and is easily confused with CXB emission using this extraction method \citep{2021MNRAS.502.3966K}.
Further, to try to account for this possible detector issue, the tying of detector values to the relative ratio difference to {\tt DET0} as estimated from the OCC NOSUN data series and then applying a maximum value helped FPMB somewhat at energies above 10~keV, but FPMA still exhibited the lower than expected aCXB normalization values. Implementing our assumptions on the effect the higher detector normalization values FPMA have in two of its detectors, we excluded all data from {\tt DET3} as it was consistently found to report higher values then all other detectors. The resulting measurement gave a greater agreement between FPMA and FPMB through all energies and confirmed that these higher than expected detector normalization values were surpressing the CXB normalization value in energies where the detector and CXB are within an order of magnitude.
The effect of this solar flux was noticeable, but did not seem to be the source of the disagreement between this study and that of the pilot study. At this time it is unknown as to the difference, as both studies used similar fitting techniques and preparation of data. With persistent values found with separation of the data by region and time, and fitting techniques that reproduce standard measurements of the Crab nebula, it is not clear as to the reason our two measurements differ, though the improvement to FPMB when applying boundaries to the detector normalization values does give credence to the need to treat the detector values carefully. 

\begin{figure}[h]
    \centering
    \includegraphics[width=\linewidth]{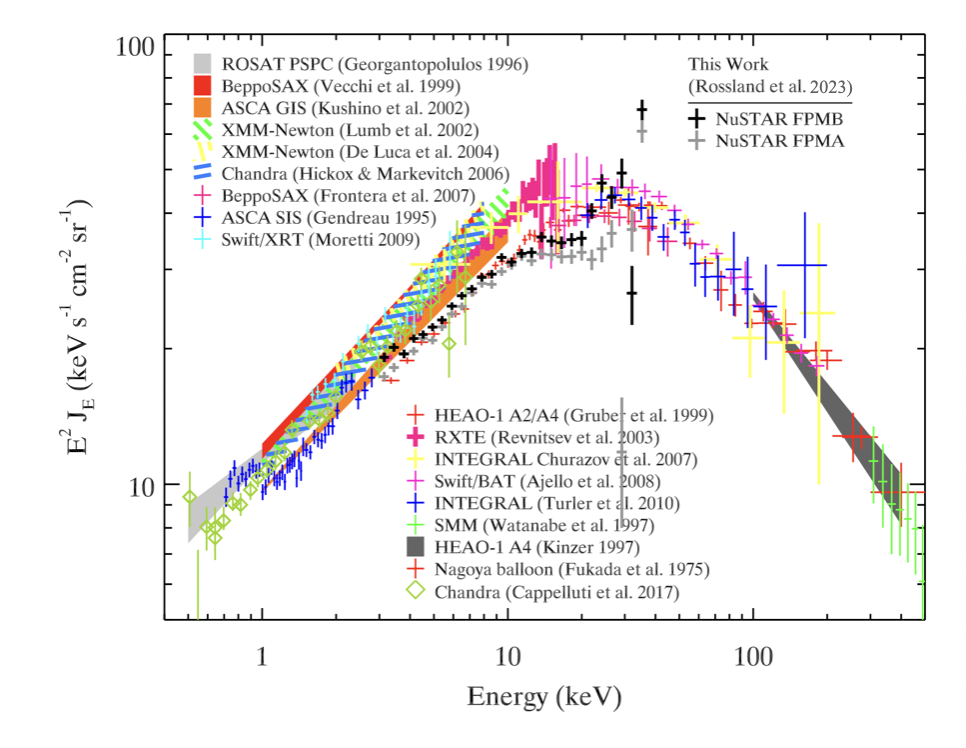}
    \caption{Same as Fig.~\ref{fig:cxb324} but including the less reliable \nustar\ measurements up to 40~keV, also placing no a priori constraints on the detector components $a_0$ through $a_3$ while including all detectors for both telescopes. Below 10~keV, the \nustar\ measurements still follow the \heao\ measurements closely, while above 10~keV both telescopes display more deviation from the expected CXB shape. 
    }
    \label{fig:cxb340}
\end{figure}

In Figure~\ref{fig:cxb324} and in Figure~\ref{fig:cxb340}, we compare our CXB measurements with those from other missions. The latter figure shows \nustar\ data up to 40~keV, the highest energy where the signal is not completely swamped by the instrumental background.
Above 10~keV in FPMA and 15~keV in FPMB, our agreement with the \heao\ degrades, likely due to a loss of signal; at $E > 15$~keV, the aCXB is no longer dominant. This is supported by the full exposure time measurement showing an increase in our agreement to the \heao\ data in these energies for both FPMA and FPMB. 
From $\sim$22--30~keV, strong instrumental lines add potential systematic uncertainties within a spectral bin depending on the relative strengths of the line between detectors, which cause our measurements to scatter by more than expected given our estimate of the statistical uncertainty.
Above 20~keV, the aperture stops start to become transparent, which allows flux from a larger solid angle of sky to reach the focal planes \citep{Madsen-straylight}.
While the spatial modulation of stray light should change somewhat, it will qualitatively have the same shape, and thus our measurements rise above the expected values from past measurements due to the additional flux.
Instead of this extra stray light originating from the CXB generally, it could be caused by a handful of bright, nearby sources such as those we screened for stray light patterns.
However, this circumstance is unlikely since high fluxes are seen in all data sets, including the limited solid angle subsamples of the continugous survey fields. 
These complications suggest that measurements of the CXB with \nustar\ at energies above 24~keV would require a more detailed characterization of instrumental background components and of the aperture stop transparency.

We discuss various and sometimes subtle ways the data could be contaminated, the most important of which is illumination of the spacecraft by the Sun.
A steep and spatially varying background component due to the Sun was recognized since the early days of the \nustar\ mission \citep{2014ApJ...792...48W}, but this study has revealed two surprising aspects of this component.
As noted in \citet{2021MNRAS.502.3966K} and discussed in Section \ref{sec:selectfit}, the gradient of the solar emission on the focal planes is similar enough to that of the aCXB that it is easily confused for the CXB at low energies.
The presumption is that these photons are reflected off the observatory structure in some way that allows them to reach the detectors through the aperture stops. This is a very different path than unobstructed stray light coming from a region 1-3$^{\circ}$ from where \nustar\ is pointed.
That the pattern in the focal planes is similar in both of these cases is neither understood nor expected.
Secondly, while the solar signal does dominate the background energies at the lowest energy range of \nustar, the overall measurement of the CXB when performed by this spatial analysis is less susceptible to this extra flux than first thought. It could be, as shown in Section \ref{sec:selectfit}, that the gradient formed from solar emission may have enough of a difference from the CXB gradient that modern fitting programs can easily separate these shapes, which is in direct conflict with the assumptions previously made. Though it is more than likely due to only being seen in the first few energy bins due to its steep slope and thus only has minimal effect on the overall shape and measurement values.

Extended emission from the Galaxy---the GRXE---can also impact the inferred CXB.
We found Fe complex line emission present in CXB spectra when data from Galactic latitudes $10^{\circ} < |b| < 30^{\circ}$~was included.
Although the signal was weak, it was present in both FPMA and FPMB and suggested continuum emission from the GRXE was also contaminating our CXB spectra.
Restricting data for CXB measurements to only exclude $|b| < 10^{\circ}$~observations, as some studies have done.

\subsection{Consequences for AGN Population Models}

At hard X-ray energies, the CXB can be almost entirely traced to emission produced during accretion onto SMBHs.
This emission is contributed by AGN of different types and over a range of redshifts, constituting the total accretion history onto SMBHs over all of cosmic time.
Thus, the spectral shape and overall flux of the CXB plays an integral role in constraining the various populations of AGN, especially that of Compton-thick (CT) AGN, due to the difficulty of detecting individual sources.
The CXB can be used to help constrain these populations, also including unobstructed and Compton-thin AGNs, by integrating their luminosity functions as a function of redshift and comparing to the CXB.
Studies using directly measured AGN populations in deep surveys are unable to reproduce the CXB, especially its peak \citep[][]{Ueda2003-Chandra+,Treister2005_AGNPOP,2007A&A...463...79G}. The missing AGN flux at higher energies can be accounted for by a ``hidden'' population of AGN whose direct emission is highly absorbed \citep[][]{Setti&Woltjer1989}.
AGN population synthesis modeling can match the CXB only if a CT population---SMBHs surrounded by obscuring gas with column densities $N_H > 10^{24}$~cm$^{-2}$---exist
\citep[][]{2007A&A...463...79G,Sazonov2008,2014ApJ...786..104U}. CT AGN contribute an important fraction by numerous population synthesis models \citep[][]{Ajello2008-swift}.
Studies of the nearby universe have shown an observed population of nearby ($z \leq 0.1$) CT AGN far below what current models predict \citep[][]{Comastri2004,DellaCeca2008,Vasudevan2013,2014ApJ...786..104U,Ricci2015}, although observational biases against CT AGN could explain this discrepancy \citep[][]{Burlon2011}.

While \heao\ was explicitly designed to measure the CXB over a broad bandpass with non-imaging proportional counter instruments,
subsequent measurements at lower energies made with focusing telescopes
have reported CXB fluxes $\sim$10--30\% systematically higher than non-focusing optics \citep[][see also Fig.~\ref{fig:cxb324}]{Revnivtsev2005--HEAO1-cxb}. These more recent measurements have caused some
studies to down-weight the \heao\ CXB measurements and develop models that explain a CXB with a higher low energy flux
\citep[][]{Ueda2003-Chandra+,Treister2005_AGNPOP,Worsley2005}. The implications of a higher normalization for the CXB affects either the number density of AGNs, 
or in their radiative efficiency
\citep[][]{Shankar2008ApJ...676..131S,Merloni2007MNRAS.381..589M,laFranca2010ApJ...718..368L,Tonima2020}. 

In this work, we have measured a CXB flux more in line with that reported from \heao\, with a 3--20~keV flux $\sim10\%$ lower than more recent measurements, implying that populations of unobstructed and Compton thin AGN are more likely to have the space density described by CXB population synthesis models that match \heao\ such as \cite{Gruber99} and \cite{Revnivtsev-2003A&A...411..329R} when corrected with values from Table 3 from \cite{Revnivtsev2005--HEAO1-cxb} to account for the cross-calibration with XMM-Newton. 
Our spectral values were consistent, within errors, to those used in \cite{2016ApJ...831..185H}, which used the population synthesis model of \cite{2015MNRAS.451.1892A} folded through the \nustar\ response function. With a lower normalization then those of previous CXB measurements, we conclude that the resolved population of obscured AGN in energies of $\sim8-24~$keV is closer to $\sim40\%$.
Unfortunately, our analysis does not allow a strong constraint on the position of the peak in the CXB spectrum, except that it lies at $> 20$~keV.
Since the peak of the CXB most constrains the space density of CT AGN populations, their total contribution remains uncertain.
A successor hard X-ray mission to \nustar, such as the proposed Probe-class {\it HEX-P} observatory, would be more easily able to directly detect the sources of the CXB near its peak and quantify this elusive population.

\acknowledgements
{This work was done with the support from the \nustar\ team and from NASA Astrophysics Data Analysis Program grant 80NSSC18K0686. R. K. acknowledges support from the Russian Science Foundation (grant 19-12-00396).

This work made use of data from the NuSTAR mission, a project led by the California Institute of Technology, managed by the Jet Propulsion Laboratory, and funded by the National Aeronautics and Space Administration.

This research has made use of data and/or software provided by the High Energy Astrophysics Science Archive Research Center (HEASARC), which is a service of the Astrophysics Science Division at NASA/GSFC and the High Energy Astrophysics Division of the Smithsonian Astrophysical Observatory. 

This research made use of Astropy---a community-developed core Python package for Astronomy (Astropy Collaboration, 2013)---, numpy---Van der Walt and Colbert 2011, Computing in Science \& Engineering 13, 22---, and scipy---a community-developed open source software for scientific computing in Python \citep{Astropy,numpy,SciPy}
}

\bibliography{sample63.bbl}

\bibliographystyle{aasjournal}



\end{document}